\documentclass[11pt,a4paper]{article}
\usepackage{jheppub}

\usepackage{amsmath,amssymb,mathtools}
\usepackage{nccmath}
\usepackage[compatibility=false]{caption} 
\usepackage{subcaption} 
\usepackage{booktabs}
\usepackage{multirow}

\newcommand{\be}{\begin{equation}}
\newcommand{\ee}{\end{equation}}
\newcommand{\bal}{\begin{aligned}}
\newcommand{\eal}{\end{aligned}}

\newcommand{\bes}{\begin{split}}
\newcommand{\ees}{\end{split}}
\def\bea{\begin{eqnarray}}
\def\eea{\end{eqnarray}}
\newcommand{\Log}{\textrm{Log}}

\newcommand{\veck}{\mathbf{k}}
\newcommand{\vecq}{\mathbf{q}}
\newcommand{\veca}{\mathbf{a}}
\newcommand{\vecb}{\mathbf{b}}
\newcommand{\vecx}{\mathbf{x}}
\newcommand{\vecy}{\mathbf{y}}
\newcommand{\sech}{\mathrm{sech}}
\renewcommand{\Re}{\mathrm{Re}}
\renewcommand{\Im}{\mathrm{Im}}
\newcommand{\MBEDIT}[1]{}

 \title{A preferred ground state for the scalar field in de Sitter space}
 
 \author[a,b]{S. Aslanbeigi} 
 \author[c]{and M. Buck}

 \affiliation[a]{\small Department of Physics and Astronomy, University of Waterloo, Waterloo ON, N2L 3G1, Canada}
 \affiliation[b]{Perimeter Institute for Theoretical Physics, 31 Caroline St. N., Waterloo ON, N2L 2Y5, Canada}
 \affiliation[c]{Theoretical Physics Group, Blackett Laboratory, Imperial College, London, SW7 2AZ, U.K.}

\abstract{ 
We investigate a recent proposal for a distinguished vacuum state of a free scalar quantum field in an arbitrarily curved spacetime, known as the Sorkin-Johnston (SJ) vacuum, by applying it to de Sitter space.
We derive the associated two-point functions on 
both the global and Poincar\'e (cosmological) patches in general $d+1$ dimensions. 
In all cases where it is defined, the SJ vacuum belongs to the family of de Sitter invariant $\alpha$-vacua. We obtain different states depending on the spacetime dimension, mass of the scalar field, and whether the state is evaluated on the global or Poincar\'e patch. We find that the SJ vacuum agrees with the Euclidean/Bunch-Davies state for heavy (``principal series'') fields on the global patch in even spacetime dimensions. We also compute the SJ vacuum on a causal set corresponding to a causal diamond in $1+1$ dimensional de Sitter space. Our simulations show that the mean of the SJ two-point function on the causal set agrees well with its expected continuum counterpart.
}
\begin{document}
\maketitle
\flushbottom
\section{Introduction}
``Quantum field theory (QFT) in curved spacetime" is primarily a framework for studying the effect of spacetime geometry on quantum fields.  It is expected to provide an adequate description of nature in situations where quantum gravity effects can be ignored, such as the post-Planckian early universe. Predictions made within this framework have led to profound insights into the interplay between matter and spacetime geometry, such as  the emission of thermal radiation by black holes~\cite{hawking1975particle}, the Unruh effect~\cite{fulling1973nonuniqueness,davies1975scalar,unruh1976notes}, and the generation of Gaussian-distributed and nearly scale-invariant random perturbations in the theory of inflation~\cite{Mukhanov1992203}.

In each of these applications, physical predictions rely heavily on the choice of a ``vacuum'' or some reasonable reference state for the quantum field. 
It is well known, however, that the unique Poincar\'e-invariant vacuum of flat space does not admit an obvious generalization to arbitrarily curved backgrounds.  
A notable exception is if the spacetime admits a timelike Killing vector $\partial_t$, for which a natural choice of vacuum is the ground state of the Hamiltonian on $t=const.$ hypersurfaces.
For non-stationary spacetimes, however, 
even a large symmetry-group does not always guarantee a unique vacuum state without further input, as exemplified by the one-parameter family of $\alpha$-vacua in de-Sitter space
~\cite{allen1985vacuum}.
Since particle states are excitations built upon the vacuum, this issue calls into question the very notion of particles.

 One perspective on this issue is that the particle interpretation need not be at the heart of quantum field theory, but rather an emergent feature in suitable situations.
 This view is realized in the framework of ``algebraic quantum field theory", where the notion of a preferred state is replaced by a distinguished \textit{class} of states, the so-called Hadamard states,
 supplemented by an assumption about a short-distance asymptotic expansion for products of quantum fields (see e.g. \cite{wald2009formulation}). 
In our view, seeking a natural construction of quantum states in curved spacetime can only be fruitful, whether or not it is logically necessary for quantum field theory as such.
As argued in \cite{Afshordi:2012jf}, for instance, one can hope to find ``natural" states for the early universe, which in turn may provide some hints into the era of quantum gravity.



A proposal has recently been made for the ``ground state of a spacetime region'', which defines in a covariant way a unique state for a free quantum field in a globally hyperbolic region of an arbitrarily curved spacetime. The proposal grew out of efforts to formulate quantum field theory on causal sets~\cite{Johnston:2008za,Johnston:2009fr}, but its formulation extends naturally to continuum spacetimes. This extension was carried out and outlined in~\cite{Afshordi:2012jf}, and put on a more rigorous algebraic footing in~\cite{Fewster:2012ew}. We shall refer to the formalism as the Sorkin-Johnston (SJ) formalism, and to the state that it defines as the SJ vacuum. 

In this paper, we apply the SJ formalism to a free massive scalar field in de Sitter space, which is a particularly interesting setting for various reasons.
Firstly, it has been shown that the SJ vacuum agrees with the ground state of the Hamiltonian in static spacetimes~\cite{Afshordi:2012jf,Afshordi:2012ez}. Because de Sitter space and its half spaces are not static (or stationary), computing the SJ vacuum thereon is not merely another ``consistency check''. 
Secondly, as demonstrated in~\cite{Afshordi:2012jf}, the SJ formalism is sensitive to the global structure of spacetime. By
evaluating it on the the full de Sitter hyperboloid as well as its Poincar\'e half space, we can investigate further its nonlocal nature. 
\MBEDIT{couple of edits in the following paragraph}
Thirdly, the SJ vacuum is, strictly speaking, only defined on \textit{bounded} regions of spacetime. One strategy to find the SJ vacuum on an unbounded region is to first compute it for a bounded globally hyperbolic subregion, and then to take the appropriate limits to recover the entire spacetime. In the case of de Sitter space, we will see that this procedure gives meaningful answers in most circumstances, but that it also fails in some cases. Fourthly, it is worth investigating whether, or in which circumstances, the SJ vacuum obeys the so-called Hadamard condition. An explicit calculation in~\cite{Fewster:2012ew} shows that the SJ vacuum is not always Hadamard. We find that for certain ranges of the scalar field mass and values of spacetime dimension, the SJ vacuum on de Sitter space is also not Hadamard.
Finally, de Sitter space is appropriate for studying potential phenomenological applications of the SJ vacuum to cosmology.

\begin{table}[t!]
\begin{center}
\begin{tabular}{cccc}
\toprule
\textbf{Patch} 						&\textbf{Spacetime Dimension}	&	\textbf{SJ state for} $\mathbf{m\geq m_*}$		& 	\textbf{SJ state for} $\mathbf{m< m_*}$				\\
\midrule
\multirow{3}{*}{\vspace{12pt}Global}		&	even			&	Euclidean					&	$\alpha$-vacuum~\eqref{eq:mavaceven}	\\
\cmidrule{2-4}
								&	odd			&	in = out					&	$\alpha$-vacuum~\eqref{eq:mavacodd}		\\
\midrule
\multirow{2}{*}{\vspace{-3pt}Poincar\'e}	&	even			&	out						&	not defined					\\
\cmidrule{2-4}
								&	odd			&	in = out					&	not defined					\\
\bottomrule
\end{tabular}
\end{center}
\caption[.]{\label{tab:SJstates} The Sorkin-Johnston vacuum in the global and Poincar\'e patches of de Sitter space. Depending on the mass $m$ of the field, the SJ vacuum corresponds to different $\alpha$-vacua (the Euclidean, in- and out- vacua are all special cases of $\alpha$-vacua and in odd spacetime dimensions the in- and out-vacua coincide). The critical mass that marks these transitions is $m_*=\frac{D-1}{2\ell}$, where $D$ is the spacetime dimension and $\ell$ is the de Sitter radius.
}
\end{table}

The construction of the SJ vacuum on causal sets is of interest for two reasons. On the one hand, causal sets may be simply regarded as Lorentz-invariant (``random lattice'') discretizations of spacetime that provide a natural ultra-violet cut-off for calculations in the continuum. In this context, the SJ formalism on a causal set may serve as a simple computational procedure for calculating the two-point function of a free scalar field in an arbitratily curved spacetime, where the continuum calculations become cumbersome.
An alternative point of view, held by some researchers in quantum gravity, is that the causal set itself is the discrete physical substratum underlying continuum spacetime. In this context, the formulation of quantum field theory on causal sets is an important step towards potential phenomenological predictions of the theory. In continuum de Sitter space, a ``natural'' class of states \textit{can} be found, so an important question is whether the SJ vacuum on the causal set agrees with one of these continuum states in the appropriate ``continuum limit''. In order to address this question, we determine the SJ vacuum on a causal set that is the discrete version of a causal diamond in $1+1$ dimensional de Sitter space. We find evidence that the mean of the discrete SJ two-point function agrees well with its expected continuum counterpart.

Before delving into technicalities, let us state our results for the SJ state in the continuum (see Table \ref{tab:SJstates} for a summary).
In the cases where the prescription gives well-defined results, the SJ vacuum always corresponds to one of the de Sitter-invariant $\alpha$-vacua. Furthermore, we find that the SJ vacuum depends on  \emph{(i)} whether the mass of the field is above or below the critical value $m_{*}=\frac{D-1}{2\ell}$ (where $\ell$ is the de Sitter radius and $D$ is the spacetime dimension), \emph{(ii)} whether it is evaluated on the complete de Sitter manifold or its Poincar\'e half-space, and \emph{(iii)} whether the spacetime dimension is even or odd.\footnote{The critical mass $m_*$ separates the so-called principal ($m\geq m_*$) and complementary ($m<m_*$) series of de Sitter representations~\cite{Gazeau:2007zz}.} For a field of mass  $m\ge m_*$ in even spacetime dimensions, the SJ state corresponds to the Euclidean vacuum on the global patch and to the out-vacuum on the Poincar\'e patch. For $m<m_*$ on the Poincar\'e patch, as well as for a discrete set of mass values below $m_*$ on the global patch, the SJ prescription cannot be applied to the entire spacetime, but only to a bounded globally hyperbolic subregion of it. Table \ref{tab:SJstates} shows a summary of these results.\MBEDIT{gave some minimal context for principal/complementary}

\section{Background and the SJ vacuum}
\subsection{Quantum Field Theory in curved spacetime}
We briefly review the quantum theory of a free
real 
scalar field $\phi(x)$ in a $D=d+1$ dimensional globally hyperbolic spacetime $(M,g_{\mu\nu})$.\footnote{We use a $-+++$ signature and natural units: $\hbar=c=1$.}
Such a spacetime admits a foliation by Cauchy surfaces $\Sigma_t$ labelled by a time coordinate $t$. 
The classical equations of motion of the field are given by the Klein-Gordon (KG) equation
\be
(\Box-m^2)  \phi(x)=0\label{eq:KG},
\ee
where $\Box=\frac{1}{\sqrt{-g}}\partial_\mu\left(\sqrt{-g}g^{\mu\nu}\partial_\nu\right)$ is the Laplace-Beltrami operator and $g$ is the determinant of the metric. The advanced and retarded Green functions $G_{R,A}(x,y)$ associated with~\eqref{eq:KG} are solutions to
\be
(\Box-m^2) G_{R,A}(x,y)=\frac{1}{\sqrt{-g}}\delta^{(D)}(x-y)\label{eq:GF},
\ee
where by definition $G_R(x,y)=0$ unless $x\succ y$ (meaning that $x$ is inside or on the future light cone of $y$) and $G_A(x,y)=0$ unless $y\succ x$. These solutions are unique when $(M,g_{\mu\nu})$ is globally hyperbolic~\cite{hawking1975large}. 
Let us also define the Klein-Gordon inner product $(\cdot\,,\cdot)$ on pairs of complex solutions to~\eqref{eq:KG}:
\be
(f,g):=i\int_{\Sigma_t} \left(\overline{f} n^\mu\nabla_\mu g-gn^\mu\nabla_\mu \overline{f} \right)d \Sigma_t,
\label{eq:KGnorm}
\ee
where bar denotes complex conjugation, $\Sigma_t$ is an arbitrary Cauchy surface in $ M$, $n^\mu$ is the future-directed unit normal to $\Sigma_t$, and $d\Sigma_t$ is the induced volume element on $\Sigma_t$. (This is a well-defined inner product because it is independent of $t$ for solutions of the Klein-Gordon equation.)

To quantize the theory, we promote $\phi(x)$ to an operator (we omit hats on operators) on a Hilbert space $\mathcal H$. As well as satisfying the KG equation, we impose on $\phi(x)$ the canonical commutation relations
\be
[ \phi(x), \phi(y)]=i\Delta(x,y),
\label{eq:commutationrelations}
\ee
where $\Delta(x,y)$ is the Pauli-Jordan function, defined as the difference between the retarded and advanced Green functions: 
\be
\Delta(x,y):=G_R(x,y)-G_A(x,y).
\ee
This is the so-called Peierls form~\cite{peierls1952commutation} of the commutation relations, which is entirely equivalent to (but more explicitly covariant than) the more commonly seen equal-time commutation relations. We expand $\phi(x)$ in terms of a set of complex solutions $\{u_{\veck}\}$ of the KG equation
\be
\phi(x)=\sum_{\veck} u_\veck(x) a_\veck+\overline{u}_\veck(x)a_\veck^\dagger,
\label{eq:modesum}
\ee
where $a_\veck$ and $a_\veck^\dagger$ are the annihilation and creation operators associated with the set $\{u_\veck\}$. They satisfy the usual commutation relations
\be
[ a_{\veck}^{}, a_{\veck'}^\dagger]=\delta_{\veck\veck'}, \qquad
[ a_{\veck}, a_{\veck'}]=[ a^\dagger_{\veck}, a^\dagger_{\veck'}]=0.
 \ee
The so-called modefunctions $\{u_\veck\}$ should be orthornormal with respect to the Klein-Gordon inner product:
\be
(u_\veck,u_\vecq)=-(\overline{u}_\veck,\overline{u}_\vecq)=\delta_{\veck\vecq}, \qquad
(u_\veck,\overline{u}_\vecq)=0.
\label{eq:KGortho}
\ee
The vacuum state $|0\rangle$ associated with this expansion is defined by the condition that $ a_{\veck}|0\rangle=0\;\forall\;\veck$. 
We will refer to the state $|0\rangle$ defined in this manner as the ``vacuum associated with the modes $\{u_\veck\}$''.

As is well-known, this construction is not unique. A different set of modes $\{u'_\veck\}$ defined by a so-called Bogoliubov transformation of the modes $u_\veck$,
\be
u'_\veck(x)=\sum_\vecq A_{\veck\vecq} u_\vecq(x)+ B_{\veck\vecq} \overline{u}_\vecq(x),
\label{eq:bogtransform}
\ee
define a different representation
\be
\phi(x)=\sum_{\veck} u'_\veck(x) a'_\veck+\overline{u}'_\veck(x){a'_\veck}^\dagger
\ee 
which is also consistent with the commutation relations~\eqref{eq:commutationrelations} so long as
\be
\begin{split}
\sum_\veck A_{\veca\veck} B_{\vecb\veck}- B_{\veca\veck} A_{\vecb\veck}&=0\\
\sum_\veck A^{}_{\veca\veck} \overline{A}_{\vecb\veck}- B^{}_{\veca\veck} \overline{B}_{\vecb\veck}&=\delta^{}_{\veca\vecb}.
\end{split}
\ee
The vacuum state $|0'\rangle$ associated with these modes, i.e. the state defined by $a'_\veck|0'\rangle=0\;\forall\;\veck$, is different from $|0\rangle$ unless $B_{\veck\vecq}=0\;\forall\;\veck,\vecq$, since otherwise $a'_\veck|0\rangle\neq0$.

The Wightman (two-point) function of the field in the state 
$|0\rangle$
 is defined as
\be
W_0(x,y):=\langle0|\phi(x)\phi(y)|0\rangle.
\ee
When $|0\rangle$ is a Gaussian state, knowledge of this function fully specifies the quantum theory, since Wick's theorem then guarantees that all field correlators reduce to polynomials in $W_0(x,y)$. We will assume that $|0\rangle$ is Gaussian, since we are dealing with a non-interacting field. 
Using the definition of the commutation relations and the Wightman function, it follows that
\be
W_0(x,y)=\frac12H_0(x,y)+\frac{i}2\Delta(x,y),
\ee
where we have defined the Hadamard function or anticommutator
\be
H_0(x,y):=2\mathrm{Re}\left[W_0(x,y)\right]=\langle0|\{\phi(x),\phi(y)\}|0\rangle.
\ee
We see that the choice of a ground state $|0\rangle$ specifies the function $H_0(x,y)$, which in turn fully encodes the state, since \emph{any} state consistent with the canonical commutation relations will have the same Pauli-Jordan function.

\subsection{The SJ vacuum}
The SJ formalism defines a unique ``ground state'' on any bounded globally hyperbolic $D=d+1$ dimensional region $(M,g_{\mu\nu})$ of spacetime, by identifying the two-point function $W_{SJ}(x,y)$ with the ``positive part'' of $i\Delta(x,y)$. Let us explain what is meant by this.
\footnote{To give a rigorous definition of the SJ vacuum, fields and propagators should really be viewed as linear operators on appropriate function spaces, as is customary in the algebraic approach to QFT. However, the mathematical and notational baggage that accompanies any rigorous treatment would cloud the matters we wish to address in this paper. We will therefore sweep aside such issues here and just mention that, when care is taken and arguments are appropriately smeared over, the construction outlined below can be put on a rigorous footing~\cite{Afshordi:2012jf,Fewster:2012ew}.}
The kernel $i\Delta(x,y)$ is both antisymmetric $i\Delta(y,x)=-i\Delta(x,y)$, and hermitian $\overline{i\Delta(y,x)}=i\Delta(x,y)$. 
This is the case because $\Delta(x,y)$ is real and in any globally hyperbolic spacetime $G_A(x,y)=G_R(y,x)$. Informally, if we think of $i\Delta(x,y)$ as a hermitian and antisymmetric matrix $[i\Delta]_{xy}$, its nonzero eigenvalues are all real and appear in pairs with equal magnitude but opposite signs. 
The SJ prescription then amounts to throwing away the negative eigenvalues and 
defining $[W_{SJ}]_{xy}$ as the positive part of $[i\Delta]_{xy}$.

With the general idea in mind, let us state the SJ prescription more carefully. Consider the space $L^2(M)$ of all square-integrable functions on $M$ with the usual inner product
\be
\langle f,g\rangle:=\int_{M}\overline{f(x)}g(x)\sqrt{-g}d^Dx
\ee
for $f,g\in L^2(M)$. We define the Pauli-Jordan operator as the integral operator whose kernel is $i$ times the Pauli-Jordan function $\Delta(x,y)$:
\be
(i\Delta f)(x)=\int_ M i\Delta(x,y)f(y) \sqrt{-g(y)}d^4y.
\label{eq:ideltadef}
\ee
Then, $i\Delta$ defines a self-adjoint operator on $L^2(M)$, meaning that $\langle i\Delta f,g\rangle=\langle f,i\Delta g\rangle$.
\footnote{
More carefully, $i\Delta$ defines a symmetric operator on a dense subset of $L^2(M)$ (smooth functions of compact support) when $M$ is bounded. In this case, it can be shown that
$i\Delta$ is bounded on $L^2(M)$ when $M$ has finite spacetime volume~\cite{Afshordi:2012jf,Fewster:2012ew}, which implies that it is self-adjoint. 
}
The spectral theorem then guarantees that $i\Delta$ has a set of real eigenvalues $\{\lambda_\veca\}$, 
as well as a complete orthonormal set of eigenvectors $\{\mathfrak u_\veca(x)\}$ which satisfy \cite{RS}
\be
i\Delta \mathfrak u_\veca=\lambda_\veca \mathfrak u_\veca,
\qquad
\lambda_\veca\in\mathbb R.
\label{eq:eigenveceq}
\ee
Since $\Delta(x,y)$ is a real function, it follows that
\be
i\Delta \mathfrak u_\veca=\lambda_\veca \mathfrak u_\veca(x) \implies i\Delta \overline{\mathfrak u}_\veca=-\lambda_\veca \overline{\mathfrak u}_\veca,
\ee
which means that the non-zero eigenvectors of $i\Delta$ come in pairs:
\be
i\Delta \mathfrak u^\pm_\veca=\pm\lambda_\veca \mathfrak u^\pm_\veca,
\ee
where by definition $\lambda_\veca>0$ and $\mathfrak u^-_\veca=\overline{\mathfrak u}^+_\veca$. Moreover, these functions are orthonormal in the $L^2(M)$ inner product:
\be
\langle \mathfrak u^\pm_{\veca\vphantom{\vecb}},\mathfrak u^\pm_{\vecb}\rangle=\delta_{\veca\vecb},
\qquad\qquad
\langle \mathfrak u^+_{\veca\vphantom{\vecb}},\mathfrak u^-_{\vecb}\rangle=0.
\label{eq:SJorthonormality}
\ee
We can now split $i\Delta(x,y)$ into a positive and negative part
\be
i\Delta(x,y )=\sum_\veca Q(x,y )-\overline{Q(x,y )},
\ee
where
\be
Q(x,y )=\sum_\veca\lambda_\veca \mathfrak u_\veca^+(x)\overline{\mathfrak u}_\veca^+(y ).
\label{eq:Q}
\ee
The SJ vacuum $|SJ\rangle$ is then defined by
\be
W_{SJ}(x,y):=Q(x,y ).
\label{eq:W}
\ee
This is a valid definition for a two-point function because $(i)$ it is positive: $\langle f,W_{SJ}f\rangle\ge0$, $(ii)$ its anti-symmetrization produces the commutator: $W_{SJ}(x,y)-W_{SJ}(y,x)=[\phi(x),\phi(y)]$, and $(iii)$ it satisfies the KG equation: $(\Box_x-m^2)W_{SJ}(x,y)=0$. That $W_{SJ}$ satisfies the KG equation follows because $\Delta(x,y)$ is the difference of two Green functions, which means it itself satisfies the KG equation $(\Box_x-m^2)\Delta(x,y)=0$. Therefore, $(\Box_x-m^2) (i\Delta f)(x)=0$ for all $f$, which implies $(\Box_x-m^2) u_\veca^+(x)=(\Box_x-m^2) (i\Delta u_\veca^+)(x)/\lambda_\veca=0$.

It follows from \eqref{eq:Q} and \eqref{eq:W} that the field operator $\phi(x)$ can be expanded as a mode sum
\be
\phi(x)=\sum_{\veca} u^{SJ}_\veca(x) a_\veca+\overline{u}^{SJ}_\veca(x)a_\veca^\dagger,
\ee
where the SJ modefunctions $\{u^{SJ}_\veca\}$ are given by
\be
u^{SJ}_\veca(x):=\sqrt{\lambda^{}_\veca}\mathfrak u^+_\veca(x).
\ee
The SJ vacuum is then defined by $a_\veca |SJ\rangle=0\;\forall\;\veca$. 

The eigenvalue problem~\eqref{eq:eigenveceq} can be reduced to a set of algebraic equations as follows. 
Given any expansion of the field in terms of a set of modes $\{u_\veck\}$, the commutator function $i\Delta(x,y )$ can 
be expressed as the mode sum
\be
i\Delta(x,y )=\sum_\veck\left[u_\veck(x)\overline{u}_\veck(y )-\overline{u}_\veck(x)u_\veck(y )\right].
\label{eq:ideltasum}
\ee
This implies that we can rewrite~\eqref{eq:eigenveceq} for an eigenfunction $u^{SJ}_\veca$ with positive eigenvalue $\lambda_\veca$ as
\be
u^{SJ}_\veca(x)=\sum_\veck A_{\veca\veck} u_\veck(x)+ B_{\veca\veck} \overline{u}_\veck(x),
\label{eq:}
\ee
where we have defined
\be
\bal
 A_{\veca\veck}&=\hphantom{-}\lambda_\veca^{-1}\langle  u_\veck, u^{SJ}_\veca \rangle,\\
 B_{\veca\veck}&=-\lambda_\veca^{-1}\langle  \overline{u}_\veck, u^{SJ}_\veca\rangle.
\eal
\ee 
As the notation is meant to indicate, these coefficients define a Bogoliubov transformation. This can be checked explicitly by acting on~\eqref{eq:} with $\langle u_\veck,\cdot\,\rangle$ and $\langle \overline{u}_\veck,\cdot\,\rangle$, which yields
\be
\begin{split}
 A_{\veca\veck}
&=\frac{1}{\lambda_\veca}\sum_\vecq A_{\veca\vecq}\langle  u_\veck,  u_\vecq \rangle+ B_{\veca\vecq}\langle  u_\veck,  \overline{u}_\vecq \rangle,\\
 B_{\veca\veck}
&=\frac{-1}{\lambda_\veca}\sum_\vecq A_{\veca\vecq}\langle  \overline{u}_\veck,  u^{}_\vecq \rangle+ B_{\veca\vecq}\langle  \overline{u}_\veck,  \overline{u}_\vecq \rangle.
\end{split}
\label{eq:alphabetarelations}
\ee
Complementing these equations with the orthonormality conditions~\eqref{eq:SJorthonormality} on the SJ modes, we find the Bogoliubov conditions
\be
\begin{split}
\sum_\veck A_{\veca\veck} B_{\vecb\veck}- B_{\veca\veck} A_{\vecb\veck}&=0\\
\sum_\veck A^{}_{\veca\veck} \overline{A}_{\vecb\veck}- B^{}_{\veca\veck} \overline{B}_{\vecb\veck}&=\delta^{}_{\veca\vecb}.
\label{eq:QuasiBogTans}
\end{split}
\ee
Finding the SJ vacuum now reduces to solving the above system of equations for $A_{\veca\veck}$ and $ B_{\veca\veck}$. 
Note that this construction is only valid in a bounded region of spacetime, since otherwise 
the inner products diverge. One strategy is to impose spatial (if necessary) and temporal cut-offs, to compute the spectrum of $i\Delta$ (which in this case is completely well-defined), and to then take the
limit as the cut-off goes to infinity. This technique has been shown to work in a variety of cases \cite{Afshordi:2012jf,Afshordi:2012ez}.
\footnote{
One exception is the case of a massless scalar field in a causal diamond in $1+1$ Minkowski space \cite{Afshordi:2012ez}, though this feature is attributed to the scale-free nature of the theory.
}
We shall see below that the method can be justified in hindsight in most cases for de Sitter space too, but that it fails in one particular instance.

\section{The SJ vacuum on de Sitter space}
In this section we compute the SJ vacuum in $D=d+1$ dimensional de Sitter space. Because the SJ formalism is sensitive to global properties of spacetime, we consider both the the full space (denoted $dS^D$), and the Poincar\'e half-space (denoted $dS^D_P$), the relevant properties of which are summarized in Appendix \ref{dSgeom}. We do not consider the static patch of de Sitter, because it is known that the SJ vacuum corresponds to the ground state of the Hamiltonian in all static spacetimes~\cite{Afshordi:2012jf}.

In order to diagonalise $i\Delta$ as described above, we need to pick an arbitrary complete set of modes $\{u_\veck\}$, in terms of which we can obtain the SJ modes. A convenient choice are the modes associated with the so-called Euclidean or Bunch-Davies (BD) state~\cite{bunch1978quantum}. The modes that define this vacuum on the full space (denoted $dS^D$) and on the Poincar\'e half-space (denoted $dS^D_P$) will be referred to as the Euclidean modes $\vspace{-1pt}u^{E}_{Lj}(x_G)$, and BD modes $u^{BD}_\veck(x_P)$, respectively, where $x_G$ and $x_P$ denote the coordinates on the two patches.
These modes define the same state on $dS^D_P$, i.e. their two-point functions are identical. 
The Euclidean/BD state belongs to a two-real-parameter family of de Sitter-invariant vacuum states, known as the \textit{Mottola-Allen} or \textit{$\alpha$}-vacua~\cite{Mottola:1984ar,allen1985vacuum}. 
We have listed the basic properties of these vacua in Appendix \ref{dSVac}.
In this section, we will show how the SJ vacuum is related to these vacua. 

\subsection{The SJ vacuum on the Poincar\'e patch}
In cosmological coordinates, the de Sitter metric reads (see Section \ref{CosmoCoord} of 
Appendix \ref{dSgeom})
\be
ds^2=\frac{\ell^2}{\eta^2}\left[-d\eta^2+\sum_{i=1}^{d}dx_{i}^2\right],
\ee
where $\eta\in(-\infty,0)$, and $x_i\in(-\infty,+\infty)$.
The positive-frequency modes that define the BD vacuum $|BD\rangle$ on $dS^D_P$ take the form
\be
 u^{BD}_\veck(\eta,\textbf{x})=\frac{e^{i\veck\cdot\vecx}}{(2\pi)^{d/2}}\chi_k(\eta),\qquad
 \chi_k(\eta)=\sqrt{\frac{\pi\ell}{4}}e^{i\pi\left(\frac{\nu}{2}-\frac{d+2}{4}\right)}\left(\frac{-\eta}{\ell}\right)^{d/2}H_{\nu}^{(1)}(-k\eta),
 \label{eq:poincaremodes}
\ee
where
\be
\nu=\ell\sqrt{m_*^2-m^2}, \qquad m_*=\frac{d}{2\ell}.
\ee
We shall refer to $m_*$ as the \emph{critical mass}. As $m$ increases from $0$ to $m_*$, $\nu$ decreases along the real line from $\frac{d}{2\ell}$ to $0$, and as $m$ increases further across $m_*$, $\nu$ moves up the imaginary axis. (The critical mass plays a central role in the representation theory of the de Sitter group; see~\cite{tagirov1973consequences,Bros:2006gs} for some considerations on the physical significance of $m_*$.)
As shown in Section \ref{AppBD} of Appendix \ref{dSVac}, these modes satisfy the KG equation and are orthonormal with respect to the KG inner product.  
The $L^2$ inner product of these modes is also of interest to us:
\bea
\langle u_\veck^{BD},u_\vecq^{BD}\rangle&=&\delta^{(d)}(\veck-\vecq)\langle\chi_k,\chi_k\rangle_{\eta}, \\
\langle u_\veck^{BD},\overline{u}_\vecq^{BD}\rangle&=&\delta^{(d)}(\veck+\vecq)\langle\chi_k,\overline{\chi}_{k}\rangle_{\eta},
\eea
where we have defined the inner product $\langle\cdot,\cdot\rangle_{\eta}$ for functions of $\eta$ only:
\be
\langle f,g\rangle_{\eta}:=\int_{\eta_{\min}}^{\eta_{max}}\overline{f(\eta)}g(\eta)\left(\frac{-\ell}{\eta}\right)^{d+1}d\eta.
\ee
We have introduced $\eta_{min}$ and $\eta_{max}$ as regulators which will be sent to $-\infty$ and $0$ (respectively) after the SJ vacuum has been computed. 
The algebraic relations \eqref{eq:alphabetarelations} and \eqref{eq:QuasiBogTans} can now be solved for:
\bea
 A_{\veck\vecq}&=&\delta^{(d)}(\veck-\vecq)\cosh(\alpha_k)\notag \\
 B_{\veck\vecq}&=&\delta^{(d)}(\veck+\vecq)\sinh(\alpha_k) e^{i\beta_k}\notag\\
 \lambda_{\veck}&=&\sqrt{\langle\chi_k,\chi_k\rangle_{\eta}^2-\left|\langle\overline{\chi}_k,\chi_k\rangle_{\eta}\right|^2},
\eea
where 
\be
\alpha_k=\frac12\mathrm{tanh}^{-1}|r_k|, \qquad
\beta_k=\mathrm{arg}(r_k)+\pi, 
\label{eq:BogGeneral}
\ee
and
\be
r_{k}:=\frac{\langle\overline{\chi}_k,\chi_k\rangle_{\eta}}{\langle\chi_k,\chi_k\rangle_{\eta}}.
\label{eq:poincareratio}
\ee
The above expressions are valid only when $|r_k|\neq1$. When $|r_k|=1$, the Bogoliubov coefficients blow up and the SJ prescription is no longer valid. 
In Appendix \ref{app:poincare}, we have computed $r_{k}$ in the limit  $\eta_{min}\to-\infty$ and $\eta_{max}\to0$:
\be
r_{k}=
\begin{dcases} 
e^{i\pi\left(\nu-\frac{d}{2}\right)} & \text{if $m\le m_*$,}
\\
e^{-i\pi\frac{d}{2}}\sech(\pi|\nu|) & \text{if $m\ge m_*$.}
\end{dcases}
\ee
We see that for masses $m\le m_*$, the SJ prescription is not well defined  in the limit $\eta_{max}\to0$, since in that case $|r_k|\rightarrow1$. When $m> m_*$, we find that the Bogoliubov coefficients are
\be
\alpha_k=\tanh^{-1}e^{-\pi|\nu|}
\quad \mathrm{and}\quad
\beta_k=-\frac{D+1}2\pi.
\ee
This corresponds to the particular $\alpha$-vacuum known as the \emph{out}-vacuum (see Section \ref{MAstates}).
More specifically, when $m>m_*$, the two point function of the SJ vacuum in the Poincar\'e patch is equal to the restriction of the out-vacuum two-point function in this region.

\subsection{The SJ vacuum on the global patch}
In global coordinates, the de Sitter metric reads (see Section \ref{GlobalCoord} of Appendix \ref{dSgeom})
\be
ds^2=-dt^2+\ell^2\cosh^2(t/\ell)\,d\Omega^2_{d},
\ee
where $d\Omega^2_{d}$ is the line element on the $d-$Sphere ($S^d$) and $t\in(-\infty,+\infty)$.
Letting $z(t)=1+e^{2t/\ell}$, the positive-frequency modes that define the Euclidean vacuum
on $dS^D$ take the form (see Section \ref{AppBD} of Appendix \ref{dSVac})
\be
u^E_{Lj}(t,\Omega)=y^E_{L}(t)Y_{Lj}(\Omega), \qquad
y^E_L(t)=\mathcal{N}_Le^{(a+\nu)t/\ell}\cosh^L(t/\ell)F(a,a+\nu;2a;z(t)-i\epsilon),
\ee
where
\be
\mathcal{N}_L=\frac{e^{i\frac{\pi}{2}(a+\nu)}}{2^a\ell^{\frac{d-1}{2}}}\frac{\sqrt{\Gamma(a+\nu)\Gamma(a-\nu)}}{\Gamma(a+\frac{1}{2})}, \qquad
a=L+d/2.
\ee
Here $F$ denotes the hypergeometric function ${}_2F_1$ and $-i\epsilon$ determines the side of the branch cut (from $1$ to $\infty$ along the real axis) where it should be evaluated. The functions $Y_{Lj}(\Omega)$ are spherical harmonics on $S^d$, whose relevant properties we have included in
Section \ref{AppBD}. Also, 
$L\in\{0,1,2,\dots\}$ and $j$ is a collective index for $j_1,j_2,\dots,j_{d-1}$, which run over values $|j_{d-1}|\le j_{d-2}\le\cdots\le j_1\le L$.
These modes satisfy the Klein-Gordon equation and are orthonormal with respect to the
Klein-Gordon inner product. 
The $L^2$ inner products of interest are
\bea
\langle u^{E}_{Lj},u^{E}_{L'j'}\rangle&=&\langle y^{E}_L,y^{E}_L\rangle_{t}\delta_{LL'}\delta_{jj'}, \\
\langle \overline{u}^{E}_{Lj},u^{E}_{L'j'}\rangle&=&\langle \overline{y}^{E}_L,y^{E}_L\rangle_{t}(-1)^L\delta_{LL'}\delta_{jj'},
\eea
where we have defined an inner product $\langle\cdot,\cdot\rangle_{t}$ for functions of $t$ only:
\be
\langle f,g\rangle_{t}=\int_{-T}^{T}\overline{f(t)}g(t)\ell^d\cosh^d(t/\ell)dt.
\ee
We have introduced $T$ as a regulator which will be sent to $\infty$ once the SJ vacuum is computed. This procedure clearly breaks de Sitter invariance, but we shall see that when the limit is taken, we obtain a state that is de Sitter invariant. The algebraic relations \eqref{eq:alphabetarelations} and \eqref{eq:QuasiBogTans} can now be solved for in complete analogy to the previous section:
\bea
 A_{Lj,L'j'}&=&\cosh(\alpha_L)\delta_{LL'}\delta_{jj'} \notag\\
 B_{Lj,L'j'}&=&\sinh(\alpha_L) e^{i\beta_L}\delta_{LL'}\delta_{jj'}\notag\\
 \lambda_{Lj}&=&\sqrt{\langle y^{E}_{L},y^{E}_{L}\rangle_t^2-\left|\langle \overline{y}^{E}_{L},y^{E}_{L}\rangle_t\right|^2},
\eea
where 
\be
\alpha_L=\frac12\mathrm{tanh}^{-1}|r_L|, \qquad
\beta_L=\mathrm{arg}(r_L)+\pi, 
\label{eq:BogGeneral2}
\ee
and
\be
r_L:=(-1)^L\frac{\langle \overline{y}^{E}_L,y^{E}_L\rangle_{t}}{\langle y^{E}_L,y^{E}_L\rangle_{t}}.
\label{eq:globalratio}
\ee
In Appendix \ref{app:global}, we have computed $r_L$ in the limit $T\to\infty$:
\be
r_L=
\begin{dcases} 
\sin\frac{D}{2}\pi\,\textrm{sech}\,\pi|\nu| & \text{if $m\geq m_*$,}
\\
\sin\left[\left(\frac{D}{2}-\nu\right)\pi\right] &\text{if $0<m\le m_*$.}
\end{dcases}\label{eq:globalcoeffs}
\ee
Regardless of the spacetime dimension or mass of the field, the SJ vacuum is invariant under the full de Sitter group in the global patch (see Appendix \ref{MAstates}).
As a result, it is always an $\alpha$-vacuum.
The case of even and odd spacetime dimensions look quite different, so we consider them in turn. For even $D$,~\eqref{eq:globalcoeffs} reduces to
\be
r_L=
\begin{cases}
0 & \text{if $m\geq m_*$,}
\\
(-i)^{D-2}\sin\pi\nu &\text{if $0<m\le m_*$,}
\end{cases}
\ee
and for odd 
$D$ we have
\be
r_L=
\begin{cases} 
(-i)^{D-1}\textrm{sech}\,\pi|\nu| & \text{if $m\geq m_*$,}
\\
(-i)^{D-1}\cos\pi\nu &\text{if $0<m\le m_*$.}
\end{cases}
\ee
When $m\geq m_*$ and $D$ is even, $\alpha_L=0$ and the SJ vacuum is equal to the Euclidean state. In odd spacetime dimensions and above the critical mass we have
\be
\alpha_L= \tanh^{-1}e^{-\pi|\nu|}
\qquad\mathrm{and}\qquad
\beta_L=-\frac{D+1}2\pi,
\ee
which means that the SJ vacuum is the \emph{in/out}-vacuum. (The $in$ and $out$-vacua are the same in odd dimensions ~\cite{Bousso:2001mw,Lagogiannis:2011st}.)
 Below the critical mass, the Bogoliubov coefficients for even $D$ are:
\be
\hspace{-35pt}\alpha_L=\frac12\tanh^{-1}|\sin\pi\nu| \quad \mathrm{and} \quad\beta_L=\left[\frac{D}{2}+\theta(-\sin(\pi\nu))\right]\pi
\label{eq:mavaceven}
\ee
and for odd $D$:
\be
\alpha_L=\frac12\tanh^{-1}|\cos\pi\nu| \quad \mathrm{and} \quad\beta_L=\left[\frac{D+1}{2}+\theta(-\cos(\pi\nu))\right]\pi,
\label{eq:mavacodd}
\ee
where $\theta(x)$ is the Heaviside step function.
In even dimensions, we obtain $\alpha=0$ whenever $|\nu|$ is an integer, in which case the SJ vacuum then corresponds to the Euclidean state. Whenever $|\nu|$ is a half-integer, the Bogoliubov coefficients diverge. The same holds in odd dimensions but with integer $\leftrightarrow$ half-integer. It is worth noting that the conformally coupled massless field corresponds in every spacetime dimension to the value $\nu=\frac12$ (through its coupling to the constant Ricci scalar, the field acquires an effective mass of $m_{cc}=\frac12\sqrt{(D-2)/(D-1)R}=\frac1{2\ell}\sqrt{D(D-2)}$, which yields $\nu=\ell\sqrt{m_*^2-m_{cc}^2}=\frac12$). Hence, the SJ vacuum for the conformally coupled massless scalar field corresponds to the Euclidean state in odd dimensions, and is ill-defined in even dimensions. A summary of the different SJ vacua in the global and Poincar\'e patches of de Sitter space is shown in Table~\ref{tab:SJstates}.\MBEDIT{Added note that v=1/2 is conformally coupled field.}\\

\begin{figure}[t]
\center
\vspace{-10pt}
\includegraphics[width=0.8\textwidth]{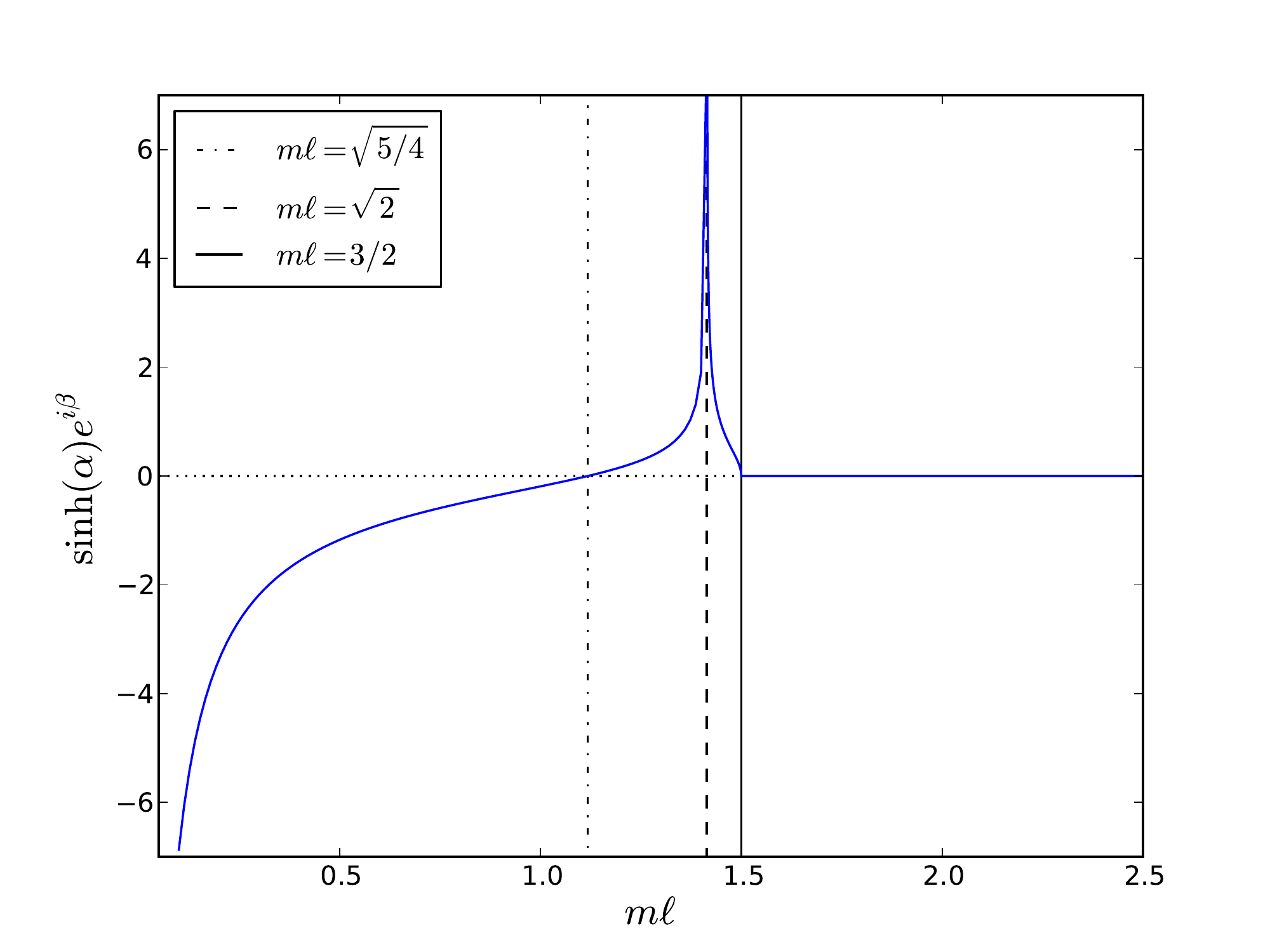}
\caption{
The Sorkin-Johnston (SJ) vacuum in the global patch of $3+1$ dimensional de Sitter space. The SJ modefunctions $u^{SJ}_{Lj}$ are related to those of the Euclidean vacuum $u^{E}_{Lj}$ by the Bogoliubov transformation $u^{SJ}_{Lj}=\cosh(\alpha)u^{E}_{Lj}+\sinh(\alpha)e^{i\beta}\overline{u}^{E}_{Lj}$, the second coefficient of which is plotted here. Depending on the product $m\ell$, where $m$ is the mass of the field and $\ell$ is the de Sitter radius, the SJ vacuum corresponds to different $\alpha$-vacua. For $m\ell\ge3/2$ and $m\ell=\sqrt{5/4}$, it coincides with the Euclidean vacuum. The prescription fails for $m\ell=\sqrt{2}$.
}
\label{fig:betaplot}
\end{figure}

Let us take a closer look at the case of macroscopic physical spacetime, $D=3+1$. As we have shown above, the SJ vacuum is the Euclidean state when $m\geq m_*=3/2\ell$. Below the critical mass, the SJ vacuum is a de Sitter invariant $\alpha$-vacuum, except when $m=m_{cc}=\sqrt{2}/\ell$, in which case the SJ prescription is not well-defined because the Bogoliubov coefficients diverge. The magnitude of the second Bogoliubov coefficient as a function of $m$ is shown in Figure~\ref{fig:betaplot}.
%


\section{The SJ vacuum on a causal set}
While the methods of canonical quantisation are not available on a causal set, the SJ formalism admits a natural discrete formulation~\cite{Johnston:2009fr,Johnston:2010su}. In fact, on a causal set, the formalism is free of many of the technicalities that arise in the continuum and accordingly simpler to outline. For the massive scalar field in $D=1+1$ dimensional flat space, it has been shown numerically that the mean of the discrete SJ two-point function approximates that of the continuum Minkowski vacuum state~\cite{Johnston:2009fr} in the appropriate ``continuum limit''. In this section, we will carry out a similar analysis in the case of two-dimensional de Sitter space.
\subsection{Causal sets and the discrete SJ vacuum}

Let us briefly review the necessary background on causal sets. A \emph{causal set} $(\mathcal C,\preceq)$ is a set $\mathcal C$ with a partial order relation $\preceq$ which is 
\be\bal
&\mathrm{(i)}&&\mathrm{reflexive:}\quad &x&\preceq x\\
&\mathrm{(ii)}&&\mathrm{acyclic:} &x&\preceq y \preceq x \implies x=y\\
&\mathrm{(iii)}&&\mathrm{transitive:} &x&\preceq y \preceq z \implies x \preceq z\\
&\mathrm{(iv)}&&\mathrm{locally\,finite:} &|\,\,&\!\!I(x,y)|<\infty
\nonumber
\eal\ee
for all  $x,y,z\in \mathcal C$, where $I(x,y):=\{z\in \mathcal C\,|\,x\preceq z\preceq y\}$ is the (inclusive) order interval between two elements $x,y\in \mathcal C$ and $|\cdot|$ denotes cardinality. We write $x\prec y$ when  $x\preceq y$ and $x\neq y$.

A causal set is fully encoded in its adjacency or \emph{causal matrix} $\mathbf C$, defined as the $|\mathcal C|\times|\mathcal C|$-matrix with entries
\be 
C_{ij} := \left\{\begin{array}{ll} 1 & \textrm{if } \nu_i \prec \nu_j \\ 0 & \textrm{otherwise,} \end{array} \right.
\label{eq:causalmatrix}
\ee 
for $\nu_i,\nu_j\in\mathcal C$, where $i,j\in \left\{1,2,\ldots,|\mathcal C|\right\}$ are indices that label the elements in $\mathcal C$.

A \emph{sprinkling} is a procedure for generating a causal set $(\mathcal C_{M},\preceq)$ given a continuum spacetime region $(M,g_{\mu\nu})$.  
Points are placed at random in $M$ using a Poisson process with ``density" $\rho:=|\mathcal C_{ M}|/V_{M}$, where $V_{ M}$ denotes the spacetime volume of $M$, in such a way that the expected number of points in any region of spacetime volume $V$ is $\rho V$. This generates a causal set whose elements are the sprinkled points, and whose partial order relation can be ``read off" from 
that of the underlying spacetime.
Such a causal set provides a discretisation of $(M,g_{\mu\nu})$ which, unlike a regular lattice, is statistically Lorentz invariant~\cite[Sec.~1.5]{henson2006causal}. In order to reduce the computational cost of the simulations described below, we keep the geodesic distance information of $(M,g_{\mu\nu})$ for all pairs of causally related elements in $\mathcal C_M$, meaning that for all such pairs $\nu_i,\nu_j\in\mathcal C_ M$ with coordinates $x_i,x_j$ in $M$, we record the values $d_{ij}:=d(x_i,x_j)$, where $d(x_i,x_j)$ denotes geodesic distance in $(M,g_{\mu\nu})$. While this information is not explicitly contained in $(\mathcal C_ M,\preceq)$, it can be recovered by known algorithms~\cite{ilie2006numerical}.

Let $(\mathcal C_M,\preceq)$ be an $N$-element causal set generated by a sprinkling into a $1+1$ dimensional spacetime region $(M,g_{\mu\nu})$. To define the SJ vacuum on the causal set, we start with the ``discrete retarded propagator'', which in two dimensions can be defined for a scalar field of mass $m$ on $\mathcal C_M$ as~\cite{Johnston:2010su} 
\be
\mathbf R=\frac12 \mathbf C\left(\mathbf 1+\frac{m^2}{2\rho}\mathbf C\right)^{-1}
\label{eq:causetretarded}
\ee
where $\mathbf C$ denotes the causal matrix defined in~\eqref{eq:causalmatrix}. It has been shown that if $(M,g_{\mu\nu})$ is a causal diamond\footnote{A causal diamond is the intersection of the interior of the past lightcone of a point $q$ with the interior of the future light cone of a point $p$ that lies to the causal past of $q$.} in two-dimensional Minkowski space, the mean of $R_{ij}$ as a function of the geodesic distance $d_{ij}$ is in agreement with the known continuum retarded propagator $G_R(x,y)$ for high sprinkling density and mass range $0<m\ll\sqrt{\rho}$~\cite{Johnston:2010su}. 
We have obtained similar evidence for the case where $(M,g_{\mu\nu})$ is a causal diamond in de Sitter space (see below). Given a retarded propagator, we define the \emph{discrete Pauli-Jordan function} $\mathbf\Delta$ on $\mathcal C_ M$ in analogy with its continuum counterpart:
\be
\mathbf\Delta:=\mathbf R-\mathbf R^T,
\ee
where $T$ denotes the matrix transpose. It is then natural to define the \emph{discrete SJ two-point function} as the positive spectral projection of $i\mathbf\Delta$:
\be
\mathbf W_{\!SJ}:=\mathrm{Pos}(i\mathbf \Delta).
\ee
Since $i\mathbf\Delta$ is now a finite Hermitian matrix (at least for causal sets of finite cardinality), its positive part is completely well-defined and specifies $\mathbf W_{\!SJ}$ uniquely. We also define the discrete analogue of the Hadamard function
\be
\mathbf H_{SJ}:=2\mathrm{Re}\mathbf W_{\!SJ},
\ee 
such that $\mathbf W_{\!SJ}= \frac12\mathbf H_{SJ}+\frac{i}{2}\mathbf \Delta$.

To compare the discrete SJ two-point function with the known propagators in continuum de Sitter space, we evaluate it on a causal set that is obtained by a sprinkling into a causal interval (diamond) in $1+1$ dimensional continuum de Sitter space. For any two points $x\prec y$, the causal interval between them is the intersection of the future of $x$ with the past of $y$. In de Sitter space, the spacetime volume $V$ of the causal interval between two timelike points depends only on their Lorentzian distance $\tau$: $V=4\ell^2\ln(\cosh(\tau\ell^{-1}/2))$. We shall refer to a causal diamond of \emph{length} $\tau$ as one whose volume is given by the formula above. 

%
\begin{figure}[t!]
\center
\includegraphics[width=0.45\textwidth]{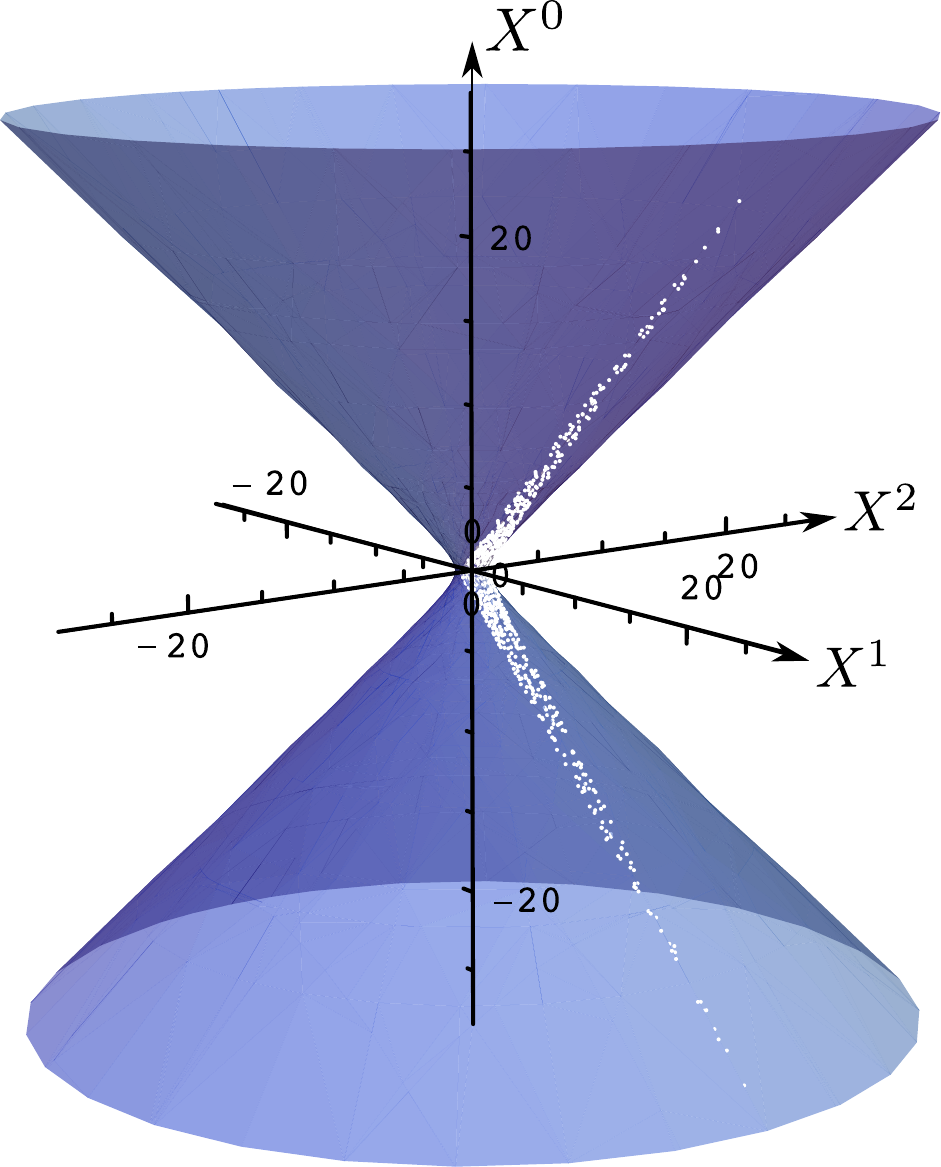}
\caption{An $N=1010$ element sprinkling with density $\rho=76\ell^{-2}$ into a causal diamond of length $\tau=8\ell$ in two-dimensional de Sitter space, visualised in the embedding three-dimensional Minkowski space (see Appendix~\ref{dSgeom}). The de Sitter radius has been set to $\ell=1$.
}
\label{fig:desittersprinkling}
\end{figure}

\subsection{Simulation results}

In order to compare causal set results with those of the continuum, we have computed the retarded propagator $\mathbf R$, and subsequently the discrete Hadamard function $\mathbf H_{SJ}$, on an $N=1010$ element sprinkling into a causal diamond of length $\tau=8\ell$ in $1+1$ dimensional de Sitter space  (implying $\rho \simeq76\ell^{-2}$). The sprinkling is shown in Figure~\ref{fig:desittersprinkling}, where we have set $\ell=1$.

\begin{figure*}[t!]
\center
\includegraphics[width=0.85\textwidth]{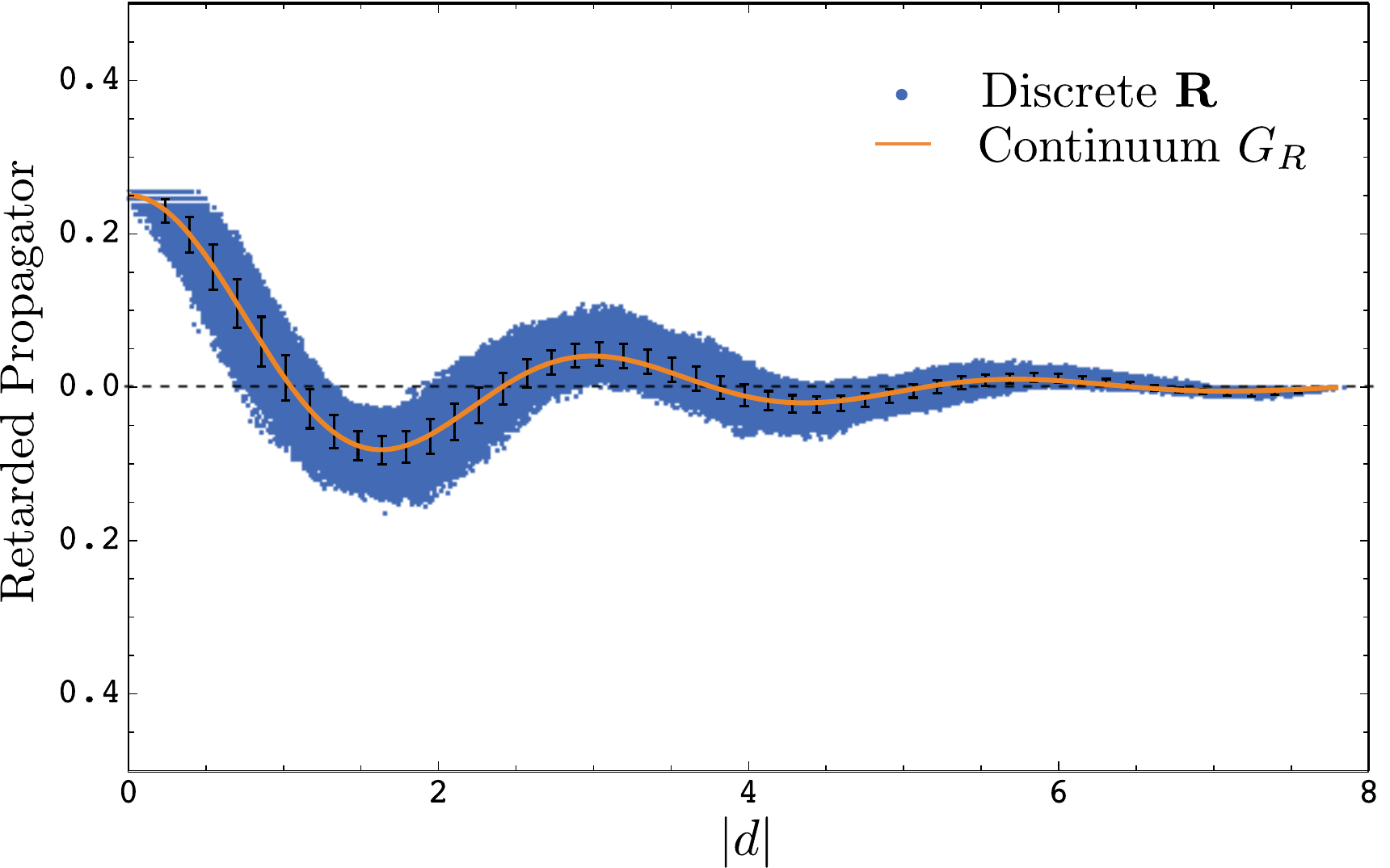}
\caption[]{
The retarded propagator $\mathbf R$, computed on a causal set obtained via a $N=1010$ sprinkling into a causal diamond of length $\tau=8\ell$ in $1+1$ dimensional de Sitter space. The mass of the field is $m=2.36\ell^{-1}$ and the de Sitter radius $\ell$ is set to unity. The geodesic distance $|d|$ between the two arguments of the function is plotted on the horizontal axis. The error bars show the standard deviation about the mean of $\mathbf R$ for binned values of $|d|$. The continuum propagator $G_R$ is shown with the thick black line.}
\label{fig:retprop}
\end{figure*}

\begin{figure}
	\begin{subfigure}[t]{0.49\hsize}
        		\includegraphics[width=\hsize]{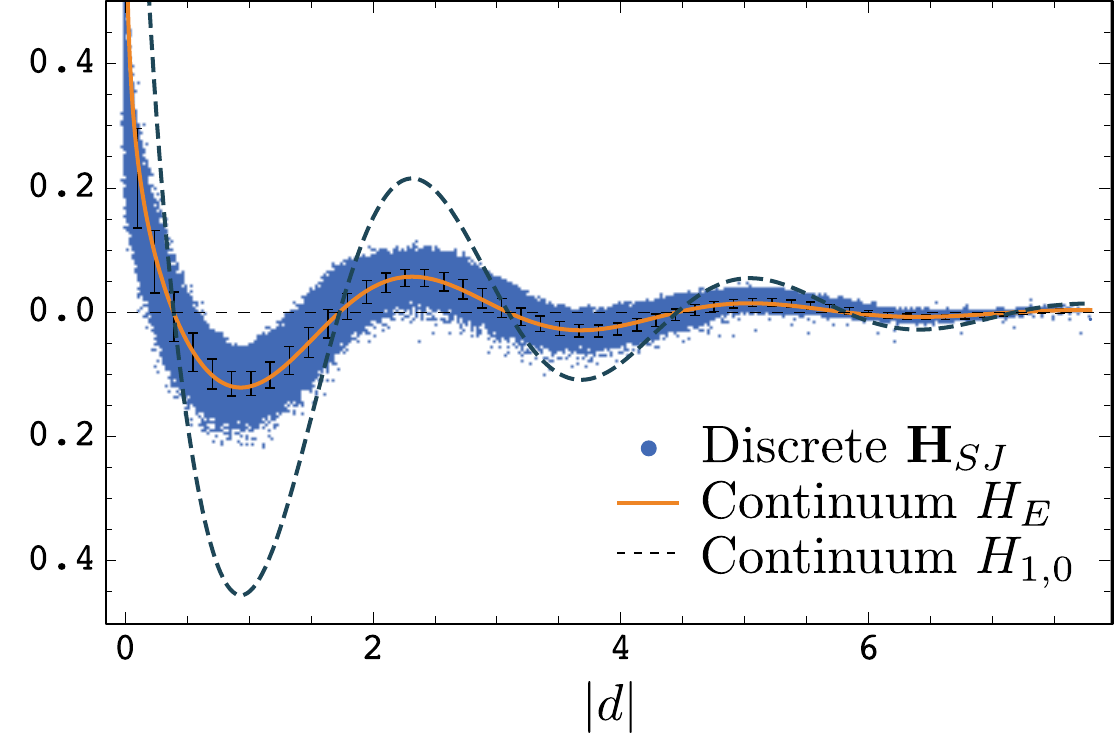}
		\caption{Timelike}	
		\label{timelike}	
    	\end{subfigure} %
	\begin{subfigure}[t]{0.49\hsize}
        		\includegraphics[width=\hsize]{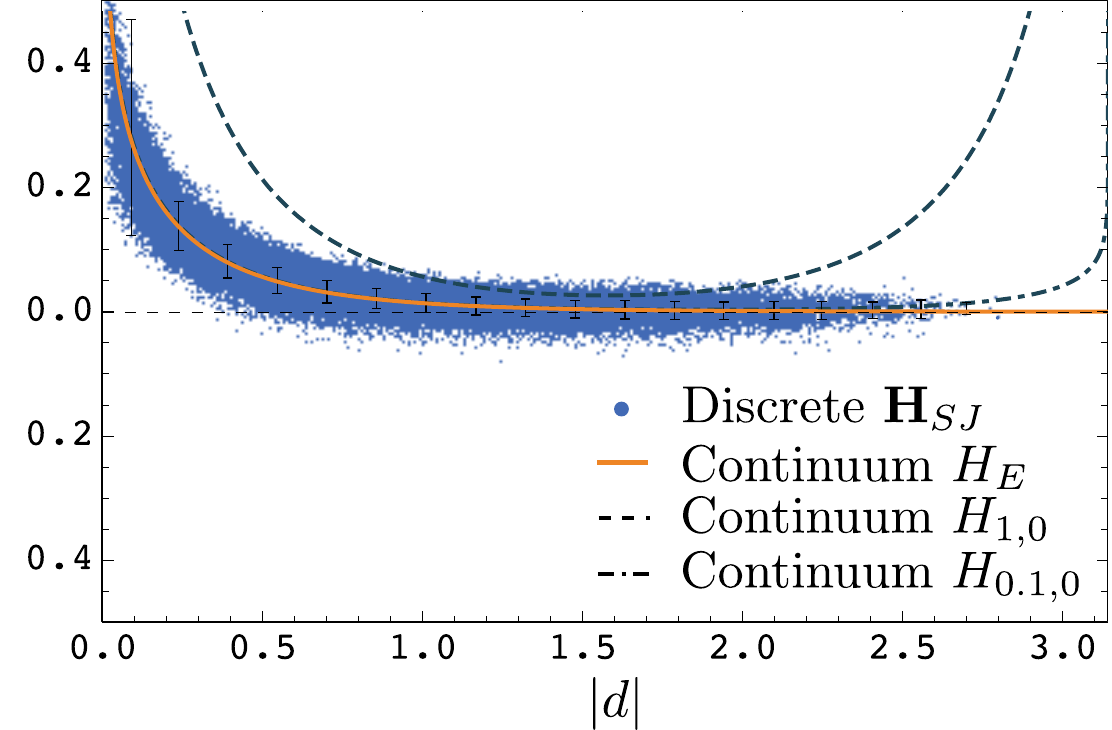}
		\caption{Spacelike}
		\label{spacelike}	
    	\end{subfigure} %
	\caption{
	The Hadamard function $\mathbf H_{SJ}$ on a causal set obtained through an $N=1010$ sprinkling of a causal diamond of length $\tau=8\ell$ in $1+1$ dimensional de Sitter space. The mass of the field is taken to be $m=2.36\ell^{-1}$, and the de Sitter radius $\ell$ is set to unity. The geodesic distance $|d|$ between the two arguments of the function is plotted on the horizontal axis for (a) timelike and (b) spacelike separated points. The error bars show the standard deviation about the mean of $\mathbf H_{SJ}$ for binned values of $|d|$. $H_{\alpha,\beta}(x,y)$ refers to the Hadamard function of the $\alpha$-vacua (see Appendix \ref{MAstates}). The function $H_{0.1,0}$ has been omitted in (a), since it is indistinguishable from the Euclidean function $H_E$.	}
	\label{fig:causetplot}
\end{figure}

Figure~\ref{fig:retprop} shows values of the retarded propagator $\mathbf R_{ij}$ for all pairs of related events $(\nu_i,\nu_j)\in\mathcal C_M$, plotted as a function their geodesic distance $d_{ij}$. 
There is good agreement between the mean of $\mathbf R$ and the continuum retarded Green function, which further validates the proposal \eqref{eq:causetretarded}. At large $\tau\gg\ell$, we see a slight deviation between the mean of the causal set data and the continuum retarded Green function. This discrepancy can be associated with edge-effects due to the finite size of the causal diamond: pairs of points separated by a geodesic distance comparable to the size of the diamond will feel the boundaries of the spacetime region (the effect of spacetime boundaries has been addressed in more detail in~\cite{Afshordi:2012ez}). 
Figure~\ref{fig:causetplot} shows the discrete SJ Hamadard function $\mathbf H_{SJ}$, computed for both timelike and spacelike pairs of events. 
Since we have no expression for the continuum SJ vacuum in the causal diamond itself, we cannot compare $\mathbf H_{SJ}$ with its \textit{exact} continuum counterpart. However, the expectation would be that the discrete SJ two-point function approximates that of a de Sitter invariant vacuum in the centre of the diamond (where the boundaries of the diamond are felt the least). Indeed, Figure~\ref{fig:causetplot} shows a very good agreement between the mean of $\mathbf H_{SJ}$ and the Hadamard function associated with the Euclidean vacuum ($\alpha=0$). At large $\tau\gg\ell$, the boundary effects become noticeable again. 
To highlight the particular agreement with the Euclidean ($\alpha=\beta=0$) Hadamard function, we have also plotted in Figure~\ref{fig:causetplot} the Hadamard function of two other $\alpha$-vacua with $(\alpha,\beta)=(1,0)$ and $(\alpha,\beta)=(0.1,0)$. Note that $H_{\alpha,\beta}(x,y)$ is more sensitive to variations in $\alpha$ for spacelike separated arguments because of the extra antipodal singularity at $d(x,y)=\pi\ell$, i.e. $Z(x,y)=-1$, present in every $\alpha$-vacuum except the Euclidean one (see Appendix~\ref{MAstates}). For instance, for the range of parameters we have probed in our simulations, including those of Figure~\ref{fig:causetplot}, the function $H_{0.1,0}$ as a function of the geodesic distance can be distinguished from the Euclidean Hadamard function for spacelike separated arguments, whereas it lies on top of the Euclidean Hadamard function for timelike separated arguments (and has thus been omitted from the timelike plot). With the parameters probed in our simulations, we cannot discriminate between the in/out and the Euclidean vacua, since they are very ``close'' unless $m\sim m_*$. Indeed, for the values presented here we have $\alpha_{in}=\alpha_{out}=\mathcal O(10^{-4})$. 
Discriminating between the in/out and Euclidean vacua is more demanding computationally. 
A full treatment of this matter will require more extensive simulations and 
is beyond the scope of our paper.

\section{Conclusions}
We have applied the Sorkin-Johnston (SJ) proposal, which defines a unique vacuum state for a free scalar field in any bounded curved spacetime, to $D=d+1$ dimensional de Sitter space.
 In those cases where the prescription is well-defined, we find that the SJ vacuum always corresponds to one of the de Sitter-invariant $\alpha$-vacua. This is reassuring, because a covariant approach should give rise to a vacuum state that respects the symmetries of the underlying spacetime. 
 We find that the SJ vacuum depends on \emph{(i)} whether the mass of the field is in the complementary or principal series (i.e. below or above the critical value $(D-1)/2\ell$),
 \emph{(ii)} whether it is evaluated on the complete de Sitter manifold or its Poincar\'e half-space, and \emph{(iii)} whether the spacetime dimension is even or odd. For a field of mass  $m<(D-1)/2\ell$ on the Poincar\'e patch, the SJ prescription cannot be applied to the entire spacetime, but only a bounded globally hyperbolic subregion of it (where the ratio~\eqref{eq:poincareratio} does not have unit modulus). It would be interesting to investigate whether a physical account can be given for the failure of the procedure in this particular case (an example of another vacuum prescription which fails for light masses is the instantaneous ground state of the Hamiltonian, particularly in the global patch~\cite{Fukuma:2013mx}). Here it is worth noting that the complementary and principal series also exhibit different behaviours in the case of interacting theories \cite{Marolf:2010zp, Jatkar:2011ju}.
For instance, quantum-corrected fields whose bare mass belong to the principal series, unlike the complementary series, decay faster than the free KG field in past/future infinity. 
This has important consequences for objects such as the S-matrix for QFTs on global de Sitter space
\cite{Marolf:2012kh,Bros:2008sq,Bros:2006gs}.

We find that the SJ vacuum in de Sitter space does not in general correspond to the
Bunch-Davies or Euclidean state, and as a result is not always Hadamard~\cite{allen1985vacuum}. (See~\cite{Fewster:2012ew} for
another instance where the SJ state is not Hadamard.) The main advantage of Hadamard
states is that for such states it is known how to construct physically relevant expectation
values, such as those of the stress-energy tensor, on arbitrarily curved spacetimes~\cite{2000CMaPh.208..623B,Hollands:2001nf, Hollands:2001fb}.
Although it has not been proven that this cannot be done for $\alpha$-vacua, it is known that
standard prescriptions such as point-splitting and normal ordering fail~\cite{Brunetti:2005pr}. We are currently exploring the consequences of these facts for the SJ formalism and hope to address them in more detail in the future.

Using the discrete SJ formalism on a causal set, we have determined the SJ state on a sprinkling of a causal diamond in $1+1$ dimensional de Sitter space. As part of our analysis, we have found evidence that the ``discrete retarded propagator'' proposed in \cite{Johnston:2008za}
agrees well with the continuum retarded propagator in de Sitter space. Our simulation also shows that the mean of the discrete SJ two-point function is consistent with that of an $\alpha$-vacuum and in particular with that of the Euclidean vacuum in the centre of the diamond (away from the edges) for a field of mass $m\ll\sqrt\rho$. This is encouraging, since the QFT defined on causal sets by the SJ formalism seems to reproduce what one would expect: a state that respects the spacetime isometries in the appropriate ``continuum limit''. It would be interesting to carry out further simulations to determine, with more statistical significance, which continuum state is best approximated by the discrete SJ state. This might be particularly illuminating when $m<m_*$, since the procedure in the continuum becomes pathological in the Poincar\'e patch in that case.


It is natural to wonder whether the SJ formalism could have phenomenological implications in relation to cosmology. We would like to raise two potential difficulties in this direction. Firstly, because of its non-local nature, it is not clear what portion of spacetime one should use to compute the SJ vacuum. For instance, should one consider the behaviour of late-time cosmology to determine the SJ vacuum for the early universe? In any case, our current calculations are not realistic because the cosmos is not \textit{always} in a de Sitter phase. It would be more interesting to compute the SJ vacuum in the case of a single-field slow-roll inflationary background, in which case the near-de Sitter phase does end. Secondly, if we ultimately aim to make a prediction for the primary anisotropy spectrum of the Cosmic Microwave Background \cite{planck}, how are we to interpret the scalar field whose vacuum state we compute using the SJ formalism? Does it also involve scalar metric perturbations? If so, one is likely to run into trouble with gauge-invariance, because the SJ formalism is not invariant under field re-definitions. We hope to address these issues in more detail in later work.

\acknowledgments
We thank Niayesh Afshordi, Dionigi Benincasa, Fay Dowker, David Rideout, Mehdi Saravani, and Rafael Sorkin for useful discussions throughout the course of this project. 
We are indebted to Ian Morrison for providing detailed comments on an earlier draft of our paper, as well as a discussion on the role of Hadamard states and the critical mass in interacting theories. 
MB thanks Perimeter Institute for hospitality. This research was supported in part by COST Action MP1006. Research at the Perimeter Institute is supported by the Government of Canada through Industry Canada and by the Province of Ontario through the Ministry of Research and Innovation.

\newpage
\appendix

\section{Geometry of de Sitter Space}
\label{dSgeom}
De Sitter space is the maximally symmetric spacetime of constant positive curvature (a comprehensive review of de Sitter geometry can be found in~\cite{schmidt1993sitter}). We denote de Sitter space in $D=d+1$ dimensions by $dS^D$. It can be viewed as the hyperboloid 
\be
X\cdot X=+\ell^2\label{eq:desitter}
\ee
in an embedding $D+1$ dimensional Minkowski space $\mathbb{M}^{D+1}$ with Cartesian coordinates $X^a$ ($a=0,1,\ldots,D$) and a Lorentzian metric $\eta_{ab}=\mathrm{diag}(-1,1,\ldots,1)$ that defines the product $X\cdot Y=\eta_{ab}X^a Y^b$. The de Sitter metric $g_{\mu\nu}$ $(\mu=0,\ldots,D-1)$ is induced by the restriction of $\eta_{ab}$ onto the hyperboloid. 

The geodesic distance between two points $p,q\in dS^D$ takes a particularly simple form in terms of the product between the embedding coordinates, which we denote by
\be
Z(p,q)
:=
\ell^{-2}X( p)\cdot X(q).
\label{ZZ}
\ee 
In terms of $Z$, the geodesic distance is 
\be
d(p,q):=\int_{\lambda_i}^{\lambda_f}\sqrt{g_{\mu\nu}\frac{dx^{\mu}}{d\lambda}\frac{dx^{\nu}}{d\lambda}}d\lambda=
\ell\cos^{-1}Z(p,q),
\ee
where $\lambda$ parametrises the geodesic $x^{\mu}(\lambda)$, and $p$ and $q$ have coordinates $x^{\mu}(\lambda_i)$ and $x^{\mu}(\lambda_f)$, respectively.
For points that can be joined by a geodesic, the range of $Z$ is $-1\le Z<\infty$, where $Z>1$, $Z=1$ and $-1\le Z<1$ correspond to timelike, null, and spacelike separations, respectively.\\

One of the symmetries of de Sitter space that will be relevant below is the antipodal map $A:p\rightarrow p^A$, which sends a point $p\in dS^D$ to its ``antipode'', denoted by $p^A$. In embedding coordinates, $A$ takes the simple form of a reflection about the origin of $\mathbb{M}^{D+1}$:
\be
X^a(p^A)=-X^a(p).
\ee 
It is clear from the invariance of~\eqref{eq:desitter} under $A$ that $p\in dS^D\iff p^A\in dS^D$. Note also that $Z(p,q)$ and $d(p,q)$ are invariant under the action of $A$.
\\

\begin{figure}[tb!]
\center
\includegraphics[width=0.5\textwidth]{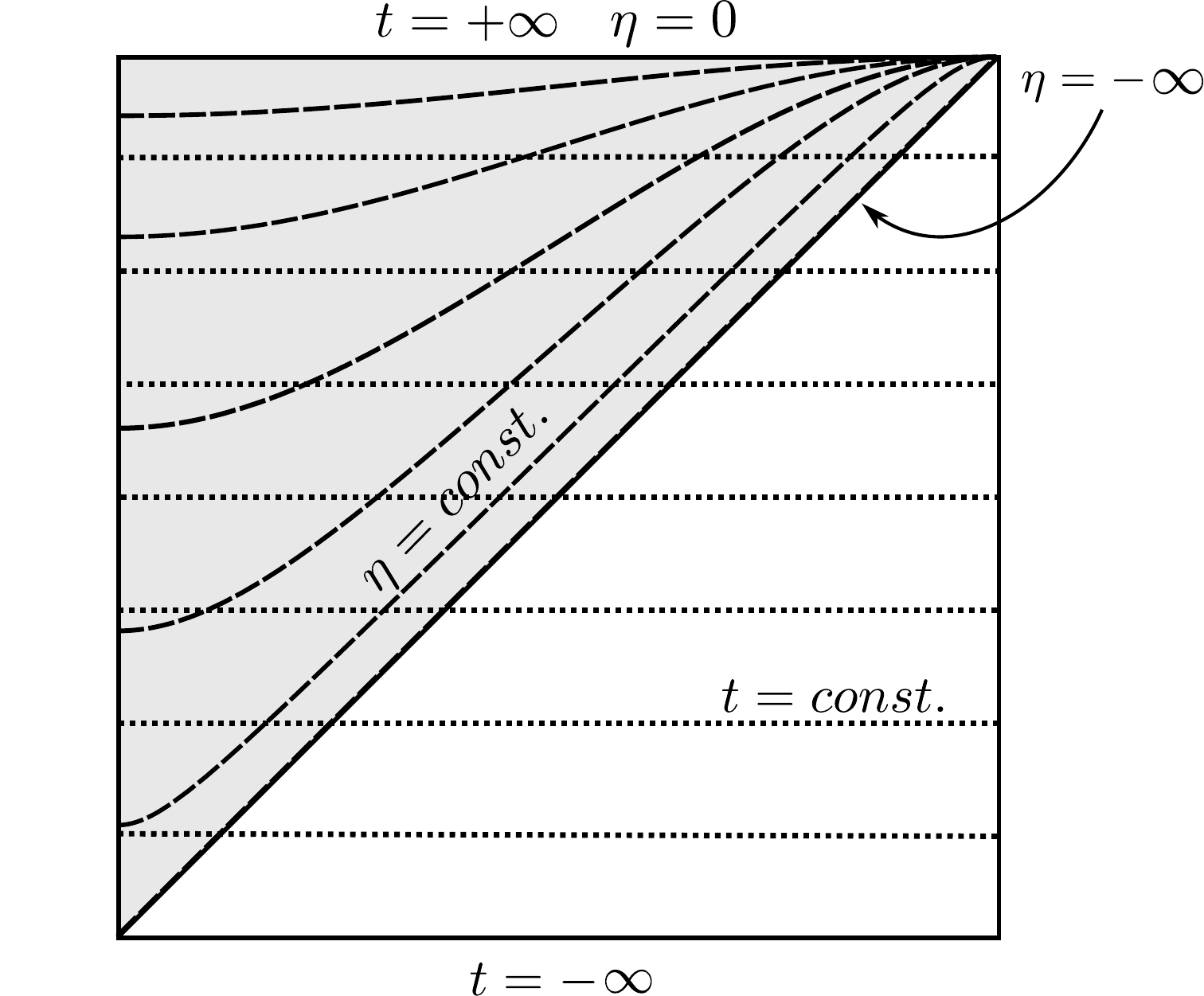}
\caption{\label{fig:desitterpenrose}The Penrose diagram of de Sitter space. The shaded area represents the (expanding) Poincar\'e patch. Dotted lines are surfaces of constant $t$ ($d$-spheres), dashed lines are surfaces of constant $\eta$ ($d$-planes). 
}
\end{figure}

We will consider two coordinate charts on de Sitter space: \emph{closed global coordinates}, which cover the entire de Sitter manifold defined by~\eqref{eq:desitter}, and \emph{cosmological coordinates}, which cover only the half space $X^0+X^1>0$, known as the (expanding) Poincar\'e patch 
(the contracting Poincar\'e patch corresponds to the other half $X^0+X^1<0$). 
We will denote the Poincar\'e patch by $dS_P^D$. It is highlighted in Figure~\ref{fig:desitterpenrose} and corresponds to the causal future of an observer at the north pole of the $d$-Sphere ($S^d$) at past timelike infinity (the bottom left corner of the Penrose diagram). De Sitter space, as well as its upper and lower half spaces, constitute globally hyperbolic manifolds in their own right, but neither admits a global timelike Killing vector field~\cite{Spradlin:2001pw} that would serve to define a unique ``minimum energy'' state.\\

\subsection{Global patch of de Sitter ($dS^D$)}
\label{GlobalCoord}
The global chart is given by the coordinates 
$x^{\mu}_G=(t,\theta^1,\ldots,\theta^d)$
defined by
\be
\begin{split}
X^0&=\ell\sinh(t/\ell)\\
X^i&=\ell\cosh(t/\ell)\,\omega^{i}\qquad\qquad 1\leq i\leq D,
\end{split}
\ee
where $\theta^i$ are the standard hyperspherical coordinates on $S^d$ and $\omega^i$ are given by
\be\bal
\omega^1&=\cos(\theta^1)\\
\omega^2&=\sin(\theta^1)\cos(\theta^2)\\
\omega^3&=\sin(\theta^1)\sin(\theta^2)\cos(\theta^3)\\
&\qquad\quad\vdots\\
\omega^{D-1}&=\sin(\theta^1)\dots\sin(\theta^{d-1})\cos(\theta^{d})\\
\omega^{D}&=\sin(\theta^1)\dots\sin(\theta^{d-1})\sin(\theta^{d}).
\eal\ee
These coordinates range over the values
\be
t\in(-\infty,\infty),\qquad
\theta^1,\dots,\theta^{d-1}\in[0,\pi],\qquad
\theta^d\in[0,2\pi).
\ee
The metric in global coordinates takes the form
\be
ds^2=-dt^2+\ell^2\cosh^2(t/\ell)\,d\Omega^2_{d},
\label{eq:globalMetric}
\ee
where $d\Omega^2_{d}$ is the line element on $S^{d}$. 
The antipode of a point $p$ with coordinates $x_G^\mu( p)=(t,\theta^1,\theta^2,\ldots,\theta^d)$ has coordinates 
$x^\mu_G(p^A)=(-t,\pi-\theta^1,\pi-\theta^2,\ldots,\pi-\theta^{d-1},\theta^d\pm\pi)$, where the $+$ and $-$ are for 
$0\le\theta^d<\pi$ and $\pi\le\theta^d<2\pi$, respectively.
\\
\subsection{Cosmological/Poincar\'e patch of de Sitter ($dS^D_P$)} 
\label{CosmoCoord}
The cosmological chart is defined by the coordinates $x^\mu_P=(\eta,\vecx)$ where $\vecx\in\mathbb R^d$ and
\be
\begin{split}
X^0&=\frac{-1}{2\eta}\left(\ell^2-\eta^2+\vecx^2\right)\\
X^1&=\frac{-1}{2\eta}\left(\ell^2+\eta^2-\vecx^2\right)\\
X^i&=\frac{-1}{\eta}x^{i-1}\qquad\qquad\qquad\qquad 2\leq i\leq D,
\label{eq:poincareCoord}
\end{split}
\ee
with $\vecx^2=\sum_{i=1}^{d}(x^i)^2$. The range of the (conformal) time coordinate is $\eta\in(-\infty,0)$, i.e. we work in the convention where time flows in the positive $\eta$-direction. 
The spatial coordinates range over the whole real line.
The line element is then given by
\be
ds^2=\frac{\ell^2}{\eta^2}\left[-d\eta^2+\sum_{i=1}^{d}dx_{i}^2\right],
\label{eq:poincareMetric}
\ee
which corresponds to an exponentially expanding Friedmann-Lema\^ itre-Robertson-Walker universe with flat spatial sections.\\

The antipodal map $A$ is not defined on $dS^D_P$: if $q$ is a point on the Poincar\'e patch, its antipode $q^A$ is \emph{not} a point on the Poincar\'e patch, since the antipodal map in cosmological coordinates takes the form $x_P^\mu(q)=(\eta,\vecx)\implies x_P^\mu(q^A)=(-\eta,\vecx)$, and $\eta$ is only defined on the negative real line. Bearing this in mind, we shall still use the notation $x_P^A$ on cosmological coordinates to mean ``switch the sign of $\eta$'' in $x_P$. 

\section{Vacuum states on de Sitter space}
\label{dSVac}
Here we review the so-called Euclidean or Bunch-Davies (BD) vacuum state for a massive free scalar field on de Sitter space.
The Euclidean/BD state belongs to a two-real-parameter family of de Sitter-invariant vacuum states, known as the \textit{Mottola-Allen} or \textit{$\alpha$}-vacua. We review below how these states are constructed and how they are related to each other. 
\subsection{Bunch-Davies modes on $dS^D_P$}
\label{AppBD}
In cosmological coordinates, the de Sitter metric is given by \eqref{eq:poincareMetric}.
Consider the mode functions
\be
u_\veck(\eta,\textbf{x})=\frac{e^{i\veck\cdot\vecx}}{(2\pi)^{d/2}}\chi_k(\eta),\qquad
\chi_k(\eta)=\mathcal{N}_{\veck}(-\eta)^{d/2}\psi_k(\eta),
\ee
where $\mathcal{N}_{\veck}$ is a normalisation constant and $k:=|\veck|$. These modes satisfy the Klein-Gordon equation if $\psi_k(\eta)$ satisfies
Bessel's differential equation:
\be
z^2\frac{d^2\psi_k}{dz^2}+z\frac{d\psi_k}{dz}+(z^2-\nu^2)\psi_k=0,
\label{bessel}
\ee
where
\be
z=-k\eta, \qquad
\nu^2=\frac{d^2}{4}-m^2\ell^2.
\label{nunu}
\ee
The BD positive-frequency modes are taken to be $\psi^{BD}_k(\eta)=H_{\nu}^{(1)}(-k\eta)$,
where $H_{\nu}^{(1)}$ is the Hankel function of the first kind. In order to fix the normalisation $\mathcal{N}_{\veck}$, we use the fact that these modes should be orthonormal with respect to the Klein-Gordon inner-product:
\be
(u^{BD}_\veck,u^{BD}_\vecq)=-(\overline{u}^{BD}_\veck,\overline{u}^{BD}_\vecq)=\delta^{(d)}(\veck-\vecq), \qquad
(u^{BD}_\veck,\overline{u}^{BD}_\vecq)=0.
\label{KGBD}
\ee
These conditions require the norm of $\mathcal{N}_{\veck}$ to be $|\mathcal{N}_{\veck}|=\sqrt{\frac{\pi}{4}}\ell^{\frac{-d+1}{2}}e^{-\pi\text{Im}(\nu)/2}$, while leaving its phase unconstrained.\footnote{
To derive this, note that in this foliation $n^{0}=\frac{-\eta}{l}$, $n^{i}=0$, and $d\Sigma=\left(\frac{-\ell}{\eta}\right)^{d}d^dx$. 
(See \eqref{eq:KGnorm} for the definition of these quantities). Then
\bea
(u^{BD}_\veck,u^{BD}_\vecq)&=&i\int\frac{e^{i(\vecq-\veck)\cdot\vecx}}{(2\pi)^{d}}\left(\frac{-\ell}{\eta}\right)^{d-1}\left[\overline{\chi}_k\partial_{\eta}\chi_q-\overline{\chi}_q\partial_{\eta}\chi_k\right]\notag\\
&=&i|\mathcal{N}_{\veck}|^2\ell^{d-1}\int\frac{e^{i(\vecq-\veck)\cdot\vecx}}{(2\pi)^{d}}(-\eta)\left[\overline{H}^{(1)}_\nu(-k\eta)\partial_{\eta}H^{(1)}_\nu(-q\eta)-H^{(1)}_\nu(-q\eta)\partial_{\eta}\overline{H}^{(1)}_\nu(-k\eta)\right].\notag
\label{temp}
\eea
Since this inner product is conserved with time, it suffices to evaluate it for $\eta\to-\infty$, where the Hankel function has the simple asymptotic form $H_{\nu}(-k\eta)\to\sqrt{\frac{-2}{\pi k\eta}}e^{-i\left(k\eta+\frac{\pi\nu}{2}+\frac{\pi}{4}\right)}$ (see $10.2.5$ of \cite{Olver:2010:NHM:1830479}). Plugging this back into the above expression, we find
\be
(u^{BD}_\veck,u^{BD}_\vecq)=\ell^{d-1}\frac{4}{\pi}e^{\pi\text{Im}(\nu)}|\mathcal{N}_{\veck}|^2\delta^{(d)}(\veck-\vecq).
\ee
The desired result now follows by requiring \eqref{KGBD}.
}
We choose the phase of $\mathcal N_\veck$ such that the mode functions satisfy the property $\overline u_\veck(x_P)=u_{-\veck}(x_P^A)$, where $x_P^A$ is the antipode of $x_P$. The function $\chi_k(\eta)$ has a branch cut that can be taken to be the negative real axis, so the more precise statement is that we require
\be
\overline u_\veck(\eta,\vecx)=u_{-\veck}(-\eta-i\epsilon,\vecx).\label{eq:bdrel}
\ee
When $\nu$ is either purely real or imaginary, $\overline{H}^{(1)}_\nu(x)=-e^{i\pi\text{Re}(\nu)}H_\nu^{(1)}(-x+i\epsilon)$ 
for real $x>0$ and small positive $\epsilon$. 
\footnote{
It follows from 10.11.9 and 10.11.5 of \cite{Olver:2010:NHM:1830479} that $H_\nu^{(1)}(-z)=-e^{-i\pi\nu}\overline{H}_{\overline{\nu}}^{(1)}(\overline{z})$. Letting $z=x-i\epsilon$, we find
$\overline{H}_\nu^{(1)}(x)=-e^{-i\pi\overline{\nu}}H_{\overline{\nu}}^{(1)}(-x+i\epsilon)$. For real $\nu$, the desired relation follows. For purely imaginary $\nu$, we get the same result by using
$H_{-\nu}^{(1)}(z)=e^{i\pi\nu}H_\nu^{(1)}(z)$ (10.4.6 of \cite{Olver:2010:NHM:1830479}).
}
Using this fact, we find that~\eqref{eq:bdrel} will be satisfied if the phase of $\mathcal N_\veck$ is $e^{i\pi\left(\frac{\text{Re}(\nu)}2-\frac{d+2}4\right)}$ and so
\be
\mathcal{N}_{\veck}=|\mathcal{N}_{\veck}|e^{i\pi\left(\frac{\text{Re}(\nu)}2+\frac{d}4\right)}=\sqrt{\frac{\pi}{4}}\ell^{\frac{-d+1}{2}}e^{i\pi\left(\frac{\nu}{2}-\frac{d+2}{4}\right)}.
\ee
Collecting our results, the positive-frequency modes that define the BD vacuum $|BD\rangle$ take the form
\be
 u^{BD}_\veck(\eta,\textbf{x})=\frac{e^{i\veck\cdot\vecx}}{(2\pi)^{d/2}}\chi_k(\eta),\qquad
 \chi_k(\eta)=\sqrt{\frac{\pi\ell}{4}}e^{i\pi\left(\frac{\nu}{2}-\frac{d+2}{4}\right)}\left(\frac{-\eta}{\ell}\right)^{d/2}H_{\nu}^{(1)}(-k\eta).
 \label{eq:poincaremodes}
\ee

It may also be verified that these modes minimise the Hamiltonian on the spatial slice at $\eta\rightarrow-\infty$.

\subsection{Euclidean modes on $dS^D$}
\label{AppEuc}
Our introduction of the Euclidean modes will follow that of \cite{Bousso:2001mw}, with some relevant additional details spelt out. 
In global coordinates, the de Sitter metric is given by \eqref{eq:globalMetric}.
Since the spatial sections are
$d$-spheres, it is natural to introduce spherical harmonics $Y_{Lj}(\Omega)$, which are a complete and orthonormal eigenbasis of the Laplacian $\nabla_{S^d}^2$ on $S^d$:
\be
\nabla_{S^d}^2Y_{Lj}=-L(L+d-1)Y_{Lj}, \qquad
\ee
\be
\sum_{Lj}Y_{Lj}(\Omega)\overline{Y}_{Lj}(\Omega')=\delta^{(d)}(\Omega,\Omega'), \qquad
\int Y_{Lj}(\Omega)\overline{Y}_{Lj}(\Omega)d\Omega=\delta_{LL'}\delta_{jj'}.
\ee
Here $L\in\{0,1,2,\dots\}$ and $j$ is a collective index for $j_1,j_2,\dots,j_{d-1}$, which run over values $|j_{d-1}|\le j_{d-2}\le\cdots\le j_1\le L$.
We work with a particular choice of harmonics $Y_{Lj}(\Omega)$ (see \cite{Bousso:2001mw}), which enjoy the useful property
\be
\overline{Y}_{Lj}(\Omega)=(-1)^LY_{Lj}(\Omega)=Y_{Lj}(\Omega^A),
\label{ccYlj}
\ee
where $\Omega^A$ is the antipodal point to $\Omega$ on $S^d$. 
Consider the modefunctions
\be
u_{Lj}(t,\Omega)=y_{L}(t)Y_{Lj}(\Omega), \qquad
y_L(t)=e^{(a+\nu)t/\ell}\cosh^L(t/\ell)v_L(t),
\ee
where $\nu$ is given by \eqref{nunu} and
\be
a=L+d/2.
\ee
These modes satisfy the Klein-Gordon equation if $v_L(t)$ is a solution to the hypergeometric differential equation
\be
z(1-z)\frac{d^2v_L}{dz^2}+\left[c-\left(a+b+1\right)z\right]\frac{dv_L}{dz}-abv_L=0,
\ee
where $c=2a$, $b=a+\nu$ and 
\be
z=z(t)=1+e^{2t/\ell}.
\ee
The Euclidean mode functions are defined by
\be
v_L(t)=\mathcal{N}_LF(a,a+\nu;2a;z(t)-i\epsilon),
\ee
where $F$ is the hypergeometric function ${}_2F_1$ and $\mathcal{N}_L$ is a normalisation constant. More precisely, $F$
stands for 
the hypergeometric function obtained by introducing a cut from $1$ to $\infty$ on the real axis. This is exactly the range of interest to us and $-i\epsilon$ determines the side of the branch cut on which the function should be evaluated. The normalisation constant $\mathcal{N}_L$ is determined by
requiring the modes to be orthonormal in the KG norm:
\be
(u_{Lj},u_{L'j'})=-(\overline{u}_{Lj},\overline{u}_{L'j'})=\delta_{LL'}\delta_{jj'}, \qquad
(\overline{u}_{Lj},u_{L'j'})=0,
\label{KGE}
\ee
which is equivalent to 
\bea
i&=&\ell^d\cosh^d(t/\ell)\left[y_L\frac{d\overline{y}_L}{dt}-\frac{dy_L}{dt}\overline{y}_L\right]\notag\\
&=&\frac{\ell^{d-1}}{2^{2a-1}}z^{2a}(z-1)^{\text{Re}(\nu)}\left\{(z-1)\left[v_L\frac{d\overline{v}_L}{dz}-\overline{v}_L\frac{dv_L}{dz}\right]-i\text{Im}(\nu)v_L\overline{v}_L\right\}.
\label{eucI}
\eea
Since the above expression is conserved in time, it suffices to look at the $z\to\infty$ (i.e. $t\rightarrow\infty$) limit. In that limit:
\bea
F(a,a+\nu;2a;z(t)-i\epsilon)&\xrightarrow{z\to\infty}&z^{-a}e^{-i\pi a}\left[\gamma+\xi e^{-\nu\ln z}e^{-i\pi\nu}\right]\label{FinInf}\\
\frac{d}{dz}F(a,a+\nu;2a;z(t)-i\epsilon)&\xrightarrow{z\to\infty}&z^{-a-1}e^{-i\pi (a+1)}\left[a\gamma+(a+\nu)\xi e^{-\nu\ln z}e^{-i\pi\nu}\right],
\eea
where all functions assume their principal values
\footnote{If $z$ and $c$ are two complex numbers, then $z^c=e^{c\Log(z)}$, where $\Log(z)=\ln(|z|)+i\Theta$, with $z=|z|e^{i\Theta}$ and $-\pi<\Theta\le\pi$.}
and
\be
\gamma=\frac{\Gamma(\nu)\Gamma(2a)}{\Gamma(a+\nu)\Gamma(a)},\qquad
\xi=\frac{\Gamma(-\nu)\Gamma(2a)}{\Gamma(a-\nu)\Gamma(a)}.
\ee
This expression is valid when $\nu\neq0,\pm1,\pm2,\dots$, $a\neq\nu$.
\footnote{
\!\!To arrive at these expressions, we have used $15.1.1$, $15.1.2$, and $15.8.2$ of \cite{Olver:2010:NHM:1830479} to obtain
\be\bal
\frac{\sin(\pi(b-a))}{\pi\Gamma(c)}F(a,b;c;z)&=&\frac{1}{\Gamma(b)\Gamma(c-a)\Gamma(a-b+1)}(-z)^{-a}F(a,a-c+1;a-b+1;1/z)\\
&+&\frac{1}{\Gamma(a)\Gamma(c-b)\Gamma(b-a+1)}(-z)^{-b}F(b,b-c+1;b-a+1;1/z).
\eal\ee
Here all functions assume their principal values, $|\text{ph}(-z)|<\pi$, and $(b-a)\neq 0,\pm1,\dots$ Then using \eqref{gamma2} to rewrite $\sin(\pi(b-a))$ in terms of
Gamma functions and \eqref{gamma1} to get $\Gamma(\pm(a-b)+1)=\pm(a-b)\Gamma(\pm(a-b))$, we find
\be
\bal
F(a,b;c;z)&=&\frac{\Gamma(b-a)\Gamma(c)}{\Gamma(b)\Gamma(c-a)}(-z)^{-a}F(a,a-c+1;a-b+1;\frac{1}{z})\\
&+&\frac{\Gamma(a-b)\Gamma(c)}{\Gamma(a)\Gamma(c-b)}(-z)^{-b}F(b,b-c+1;b-a+1;\frac{1}{z}).
\eal
\ee
We can also relate the derivative of $F$ to another hypergeometric function using $15.5.1$ of \cite{Olver:2010:NHM:1830479}: 
\be
\frac{d}{dz}F(a,b;c;z)=\frac{ab}{c}F(a+1,b+1;c+1;z).
\ee
Noting that for any complex number $c$ and $1<z<\infty$ we have $(z-i\epsilon)^{c}=e^{c\ln z}e^{ic\pi}$, and also using the fact
that $F(a,b;c;0)=1$, the desired expressions follow. 
}
Note that because $\Gamma(\overline{z})=\overline{\Gamma}(z)$, both $\gamma$ and $\xi$ are real when $\nu$ is real, and 
$\overline{\gamma}=\xi$ when $\nu$ is purely imaginary. Using these facts, evaluating \eqref{eucI} in the limit $z\to\infty$ constrains the norm of $\mathcal{N}_L$ to:
\footnote{Here we have used $15.5.5$ of \cite{Olver:2010:NHM:1830479} to rewrite $\Gamma(2a)=\pi^{-1/2}2^{2a-1}\Gamma(a)\Gamma(a+1/2)$.}
\be
|\mathcal{N}_L|^2=\frac{e^{-\pi\text{Im}(\nu)}}{2^{2a}\ell^{d-1}}\frac{\Gamma(a+\nu)\Gamma(a-\nu)}{\Gamma(a+\frac{1}{2})^2}.
\ee
Although the derivation of this result uses relations which are only valid for $\nu\ne0,1,2,\dots$, the final result is completely well-defined for such
values. Therefore, we could imagine a limiting procedure in which we add a tiny amount $\epsilon$ to an integer value of $\nu$, go through the same derivation, and then let $\epsilon$ go to zero. 

We use the freedom in the phase of $\mathcal N_L$ to choose mode functions with the useful property
\be
u_{Lj}(x_G^A)=\overline{u}_{Lj}(x_G).\label{eq:allentrickeuclidean}
\ee
Given that we have chosen spherical harmonics with the property
$\overline{Y}_{Lj}(\Omega)=Y_{Lj}(\Omega^A)$, this condition reduces to
\be
y_L(-t)=\overline{y}_L(t),
\label{minusCC}
\ee
which can be achieved by setting
\footnote{
\!To see this, let $\mathcal{N}_L=|\mathcal{N}_L|e^{i\Theta}$. It follows from the definition of $F$
(see e.g. $15.2.1$ of \cite{Olver:2010:NHM:1830479}) and $\Gamma(\overline{z})=\overline{\Gamma}(z)$:
\begin{equation}
 \overline{F}(a,a+\nu;2a;z-i\epsilon)=
    \begin{cases}
      F(a,a+\nu;2a;z+i\epsilon) \qquad \nu\text{ real} \\
      F(a,a-\nu;2a;z+i\epsilon) \qquad \nu\text{ imaginary}.\\
    \end{cases}
\label{ccHyper}
\end{equation}
Using $15.8.1$ of \cite{Olver:2010:NHM:1830479}, it may be checked that 
\bea
F(a,a+\nu;2a;z(t)+i\epsilon)&=&(1-z(t)-i\epsilon)^{-a-\nu}F(a,a+\nu;2a;z(t)/\left(z(t)-1\right)-i\epsilon)\\
&=&e^{-2(a+\nu)t/l}e^{i\pi(a+\nu)}F(a,a+\nu;2a;z(-t)-i\epsilon).
\eea
Using the relations above when $\nu$ is real, it follows from the definition of $y_L(t)$ that $y_L(-t)=e^{2i\Theta}e^{-i\pi(a+\nu)}\overline{y}_L(t)$.
The same formula in \cite{Olver:2010:NHM:1830479} also guarantees
\bea
F(a,a-\nu;2a;z(t)+i\epsilon)&=&(1-z(t)-i\epsilon)^{-a}F(a,a+\nu;2a;z(t)/\left(z(t)-1\right)-i\epsilon)\\
&=&e^{-2at/l}e^{i\pi a}F(a,a+\nu;2a;z(-t)-i\epsilon).
\eea
Using this expression and \eqref{ccHyper} when $\nu$ is purely imaginary, it follows that $y_L(-t)=e^{2i\Theta}e^{-i\pi a}\overline{y}_L(t)$. Combining these results we find
$\Theta=\frac{\pi}{2}\left[a+\text{Re}(\nu)\right]$.
}
\be
\mathcal{N}_L=|\mathcal{N}_L|e^{i\frac{\pi}{2}\left[a+\text{Re}(\nu)\right]}.
\ee
Collecting our results, the Euclidean modes are
\be
u^E_{Lj}(t,\Omega)=y^E_{L}(t)Y_{Lj}(\Omega), \qquad
y^E_L(t)=\mathcal{N}_Le^{(a+\nu)t/\ell}\cosh^L(t/\ell)F(a,a+\nu;2a;z(t)-i\epsilon),
\label{eq:globalmodes}
\ee
where $z(t)=1+e^{2t/\ell}$, $a=L+d/2$ and
\be
\mathcal{N}_L=\frac{e^{i\frac{\pi}{2}(a+\nu)}}{2^a\ell^{\frac{d-1}{2}}}\frac{\sqrt{\Gamma(a+\nu)\Gamma(a-\nu)}}{\Gamma(a+\frac{1}{2})}.
\ee

\subsection{Two-point functions and $\alpha$-vacua} 
\label{MAstates}
That the Euclidean and the BD modes define the same physical state is made apparent by the fact that the two-point function $W_E$ associated with the Euclidean modes \eqref{eq:globalmodes}, when restricted to the Poincar\'e patch, coincides with the two-point function $W_{BD}$ associated with the Bunch-Davies modes~\eqref{eq:poincaremodes}. 
(They are functions of the geodesic distance and the causal relation between their arguments, which are both coordinate independent quantities~\cite{bunch1978quantum,Tagirov:1972vv}.) For the Euclidean state,  the two-point function is given by~\cite{Tagirov:1972vv,Candelas:1976jv}
\be
\bal
W_E(x,&y)=\frac{\Gamma[h_+]\Gamma[h_-]}{4\pi\ell^2\Gamma[\frac{D}{2}]}
{}_2F_1\left(h_+,h_-,\frac{D}{2};\frac{1+Z(x,y)+i\epsilon\,\text{sign}(x^0-y^0)}{2}\right),
\label{eq:euclideantwopointfunction}
\eal
\ee
where $h_\pm=\frac{d}{2}\pm\nu$ and ${}_2F_1(a,b,c;z)$ is the hypergeometric function (see \eqref{ZZ} and \eqref{nunu} for the definitions of $Z$ and $\nu$). The $i\epsilon$ prescription selects the side of the branch cut from $Z=1$ to $Z=+\infty$ on which the function should be evaluated when $x$ and $y$ are causally related (when $x$ and $y$ are spacelike, then $Z<1$ and the values of the function below and above the real line coincide). The Hadamard function is equal to the real part $H_E(x,y)=\text{Re}\left[W_E(x,y)\right]$, which depends only on the coordinate-independent quantity $Z(x,y)$. The Pauli-Jordan function and the retarded Green function can be written in terms of $W_E(x,y)$, since $i\Delta(x,y)=2\text{Im}\left[W_E(x,y)\right]$ and $G_R(x,y)=\theta(x^0-y^0)\Delta(x,y)$.\\

We denote the two-real-parameter family of $dS$-invariant $\alpha$-vacua by $|\alpha,\beta\rangle$. Their modefunctions can be obtained through a Bogoliubov transformation~\cite{allen1985vacuum}
\be
u^{(\alpha,\beta)}_\veck=\cosh(\alpha)u^{BD}_\veck+\sinh(\alpha)e^{i\beta}\overline{u}^{BD}_{-\veck},
\label{eq:mamodesP}
\ee
for the BD modes,
and
\be
u^{(\alpha,\beta)}_{Lj}=\cosh(\alpha)u^{E}_{Lj}+\sinh(\alpha)e^{i\beta}\overline{u}^{E}_{Lj},
\label{eq:mamodes}
\ee
for the Euclidean modes. 
Here $\alpha\in\mathbb R^+$ and $\beta\in\mathbb R$ is defined modulo $2\pi$. Recall the relations between negative frequency modes and positive frequency modes taking antipodal arguments, which can be obtained for both the Euclidean modes~\eqref{eq:allentrickeuclidean}~\cite{allen1985vacuum}, and the Poincar\'e modes~\eqref{eq:bdrel} by appropriate choices of the arbitrary complex phases in the normalisation factors. Because of these relations, it is possible to express the two-point function $W_{\alpha,\beta}(x,y)$ associated to an arbitrary $\alpha$-vacuum in terms of the Euclidean/BD two-point function $W_E(x,y)$~\eqref{eq:euclideantwopointfunction}. The imaginary part of $W_{\alpha,\beta}(x,y)$ is always equal to $i\Delta(x,y)$ and hence identical for all $\alpha$-vacua. The real part, i.e. the Hadamard function, depends on $\alpha$ and $\beta$. By computing the mode sums using the $\alpha$-modes, the family of de Sitter invariant Hadamard functions $H_{\alpha,\beta}(x,y)$ can be obtained and reads~\cite{allen1985vacuum}:
\be
\bal
H_{\alpha,\beta}&(x,y )=\cosh2\alpha\,H_E(x,y )
+\sinh2\alpha\left[\cos\beta\,H_E(x^A,y )-\sin\beta \Delta (x^A,y)\right].
\label{eq:MApropagator}
\eal
\ee
The two-point function for an $\alpha$-vacuum is thus given by $W_{\alpha,\beta}(x,y)=\frac12 H_{\alpha,\beta}(x,y)+\frac{i}2\Delta(x,y)$. 
In this particular parametrisation of the $\alpha$-vacua~\cite{allen1985vacuum}, the Euclidean state corresponds to $\alpha=0$.\footnote{The relation between the parametrisation used here and that of \cite{Mottola:1984ar, Bousso:2001mw}, which uses a single complex parameter $\tilde\alpha$, is $\Re(\tilde\alpha)=\ln\tanh\alpha$ and $\Im(\tilde\alpha)=\beta$. The notation used here will be more convenient in the analysis of the SJ vacuum on a causal set, because the Euclidean state then corresponds to a finite value $\alpha=0$ instead of $\tilde \alpha=-\infty$.} 
The derivation of \eqref{eq:MApropagator} for modes on the Poincar\'e patch requires evaluating the BD Hadamard function outside its domain of validity. Specifically, one uses the property that
\bea
H^{BD}(\eta_x,\vecx; -\eta_y-i\epsilon,\vecy)&:=&\int d^d\veck \left[u^{BD}_\veck(\eta_x,\vecx)\overline{u}^{BD}_\veck(-\eta_y-i\epsilon,\vecy)+\overline{u}^{BD}_\veck(\eta_x,\vecx)u^{BD}_\veck(-\eta_y-i\epsilon,\vecy)\right]\notag\\
&=&H^{E}(x,y^A),
\eea
where $H^{E}(x,y^A)$ is the Hadamard function of the Euclidean vacuum, which is of course defined on all of de Sitter space. This implies that for a given choice of $\alpha$ and $\beta$, the two-point function associated with the modes \eqref{eq:mamodesP} is the restriction of the global $\alpha$-vacua two-point function, defined via the modes \eqref{eq:mamodes}, to the Poincar\'e patch.


Two $\alpha$-vacua which will be of special interest to us are the $in$- and $out$-vacua~\cite{Mottola:1984ar,Bousso:2001mw} :
\be
\alpha_{in}=\alpha_{out}= \tanh^{-1}e^{-\pi|\nu|}, \qquad
\beta_{in}=-\beta_{out}=\frac{D+1}2\pi,
\ee
which have no incoming/outgoing particles at past/future infinity, respectively.
\footnote{
The modefunctions associated with these choices of $\alpha$ and $\beta$ correspond to $\tilde{\phi}_{Lj}^{in}$ and $\tilde{\phi}_{Lj}^{out}$ defined in \cite{Bousso:2001mw}, which differ from the usually defined in/out modes by a constant phase. Of course, these two choices define the same vacuum state because the two-point function is insensitive to any constant-phase rescaling of modefunctions.
}
In other words, they minimise the Hamiltonian on spatial slices at $t\rightarrow\pm\infty$ in global coordinates, as shown in Figure~\ref{fig:desitterpenrose}. Notice that in odd spacetime dimensions, the $in$ and $out$-vacua are the same, i.e. they are related by a trivial Bogoliubov transformation, since then $\exp(i\beta_{in})=\exp(i\beta_{out})$ (``odd-dimensional de Sitter space is transparent''~\cite{Bousso:2001mw,Lagogiannis:2011st}). It is also worth pointing out that for masses much larger than the Hubble radius, $m\gg m_*=(D-1)/2\ell$, the $in/out$ states are ``exponentially close'' to the Euclidean state, since then $|\nu|=\frac12\ell\sqrt{m^2-m_*^2}\gg1$ and $\sinh(\alpha)\sim e^{-\pi|\nu|}$.
 

\section{Calculation of Inner Products}
\subsection{Poincar\'e chart}
\label{app:poincare}
\begin{figure}[h!]
\centering
\includegraphics[trim=20 10 20 10,clip=true,width=0.55\linewidth]{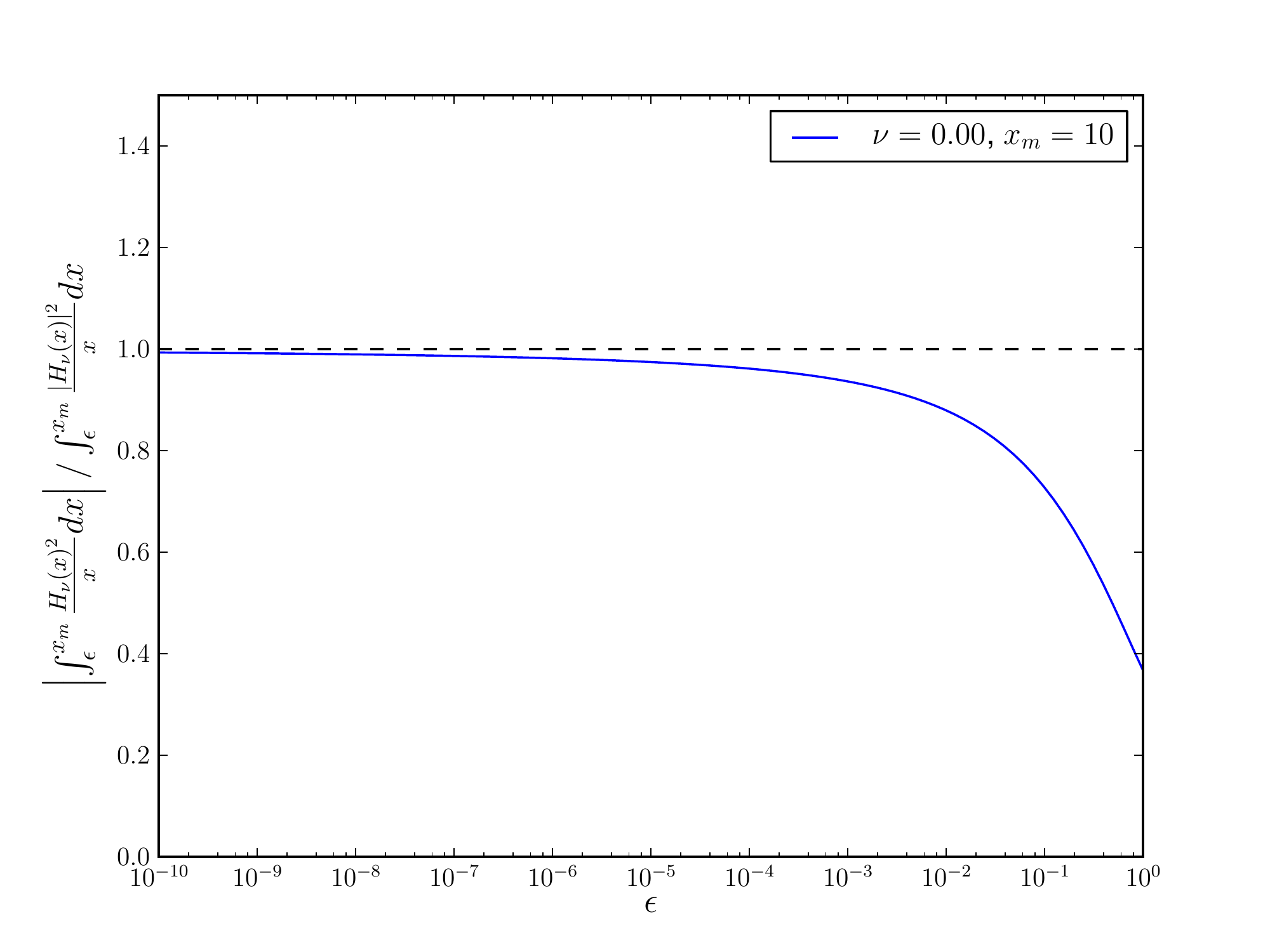}
\includegraphics[trim=20 10 20 10,clip=true,width=0.55\linewidth]{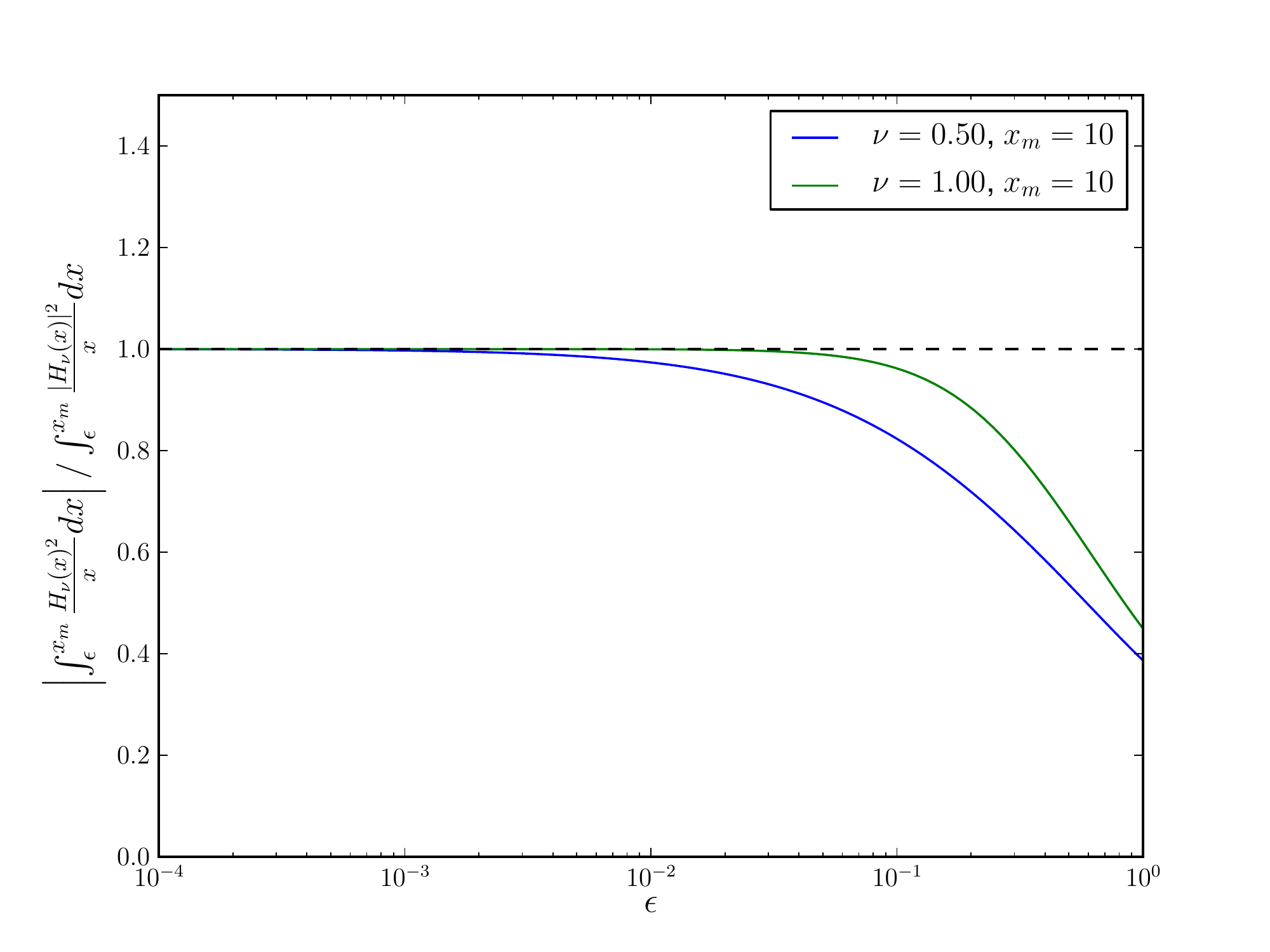}
\includegraphics[trim=20 10 20 10,width=0.55\linewidth]{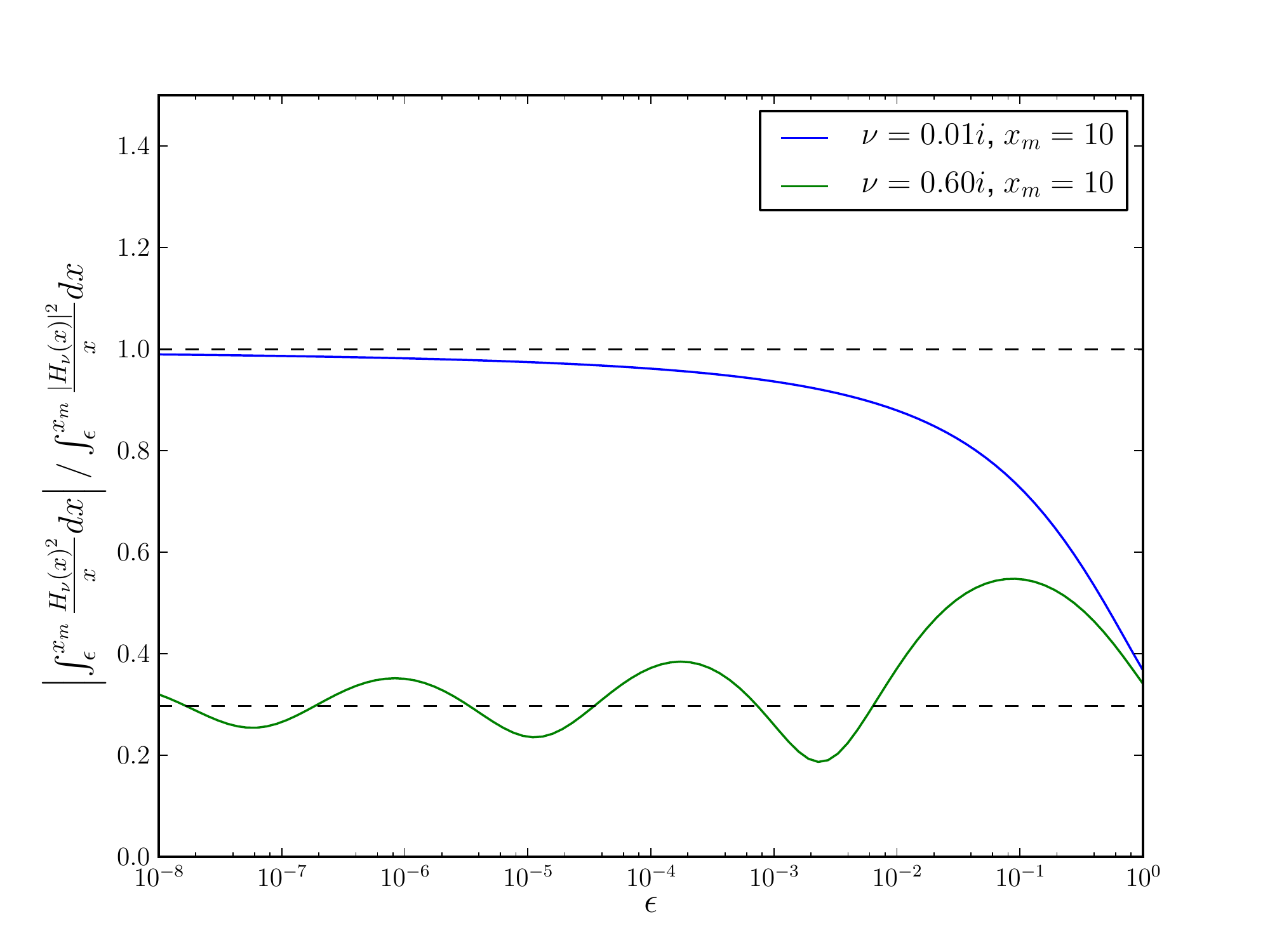}
\caption{Behaviour of $|F(\epsilon,x_m)|$ as a function of $\epsilon$. The dependence on $x_m$ is negligible for small $\epsilon$.}
\label{figrk1}
\end{figure}

Here we shall evaluate \eqref{eq:poincareratio}:
\be
r_{k}=\frac{\langle\overline{\chi}_k,\chi_k\rangle_{\eta}}{\langle\chi_k,\chi_k\rangle_{\eta}}.
\ee
It follows from the definition of $\chi_k^{BD}$ that
\bea
\langle\overline{\chi}_k,\chi_k\rangle_{\eta}&=&\frac{\pi\ell}{4}e^{i\pi\left(\nu-\frac{d+2}{2}\right)}\int_{\eta_{min}}^{\eta_{max}}\left[H^{(1)}_{\nu}(-k\eta)\right]^2\left(\frac{-\ell}{\eta}\right)d\eta\\
\langle\chi_k,\chi_k\rangle_{\eta}&=&\frac{\pi\ell}{4}e^{-\pi\Im(\nu)}\int_{\eta_{min}}^{\eta_{max}}\left|H^{(1)}_{\nu}(-k\eta)\right|^2\left(\frac{-\ell}{\eta}\right)d\eta.
\eea
Changing integration variables to $x=-k\eta$, and defining $\epsilon=-k\eta_{min}$, $x_m=-k\eta_{max}$, we find:
\be
r_{k}=e^{i\pi\left(\Re(\nu)-\frac{d+2}{2}\right)}F(\epsilon,x_m)\qquad\text{where}\qquad
F(\epsilon,x_m)=\frac{\int_{\epsilon}^{x_m}\frac{dx}{x}\left[H^{(1)}_{\nu}(x)\right]^2}{\int_{\epsilon}^{x_m}\frac{dx}{x}\left|H^{(1)}_{\nu}(x)\right|^2}.
\label{eq:poincareratio2}
\ee
Let us list a few useful properties of the Hankel function $H^{(1)}_{\nu}(z)$. It satisfies the Bessel equation $[z^2\frac{d^2}{dz^2}+z\frac{d}{dz}+(z^2-\nu^2)]H^{(1)}_{\nu}(z)=0$ and has the defining property (see $10.2.5$ of \cite{Olver:2010:NHM:1830479})
\be
H^{(1)}_{\nu}(z)\to\sqrt{\frac{2}{\pi z}}e^{i\left(z-\frac{\pi\nu}{2}-\frac{\pi}{4}\right)},
\label{eq:hank1}
\ee
as $z\to\infty$ in $-\pi+\delta\le\text{ph}z\le 2\pi-\delta$, where $\delta$ is an arbitrary small positive number. It has a branch point at $z=0$ and its principal branch corresponds to the principal value of the square root in \eqref{eq:hank1}, with a branch cut along $(-\infty,0]$.
\footnote{
\!\!$\text{PV}(z^{-\frac{1}{2}})=e^{-\frac{1}{2}\Log(z)}$, where $\Log(z)=\ln(r)+i\Theta$ with $z=re^{i\Theta}$ and $-\pi<\Theta\le\pi$.}
From here on out $H^{(1)}_{\nu}(z)$ will denote the principal value of this function. 
The asymptotic behaviour of $H^{(1)}_{\nu}(z)$ as $z\to0$ is also of interest to us:\!\!
\footnote{In~\cite{Olver:2010:NHM:1830479} see $10.7.2$ for \eqref{eq:hank2}, $10.7.7$ for \eqref{eq:hank3}, and
a combination of $10.4.3$, $10.7.3$, and $10.7.6$ for \eqref{eq:hank2}.} 
\bea
H^{(1)}_{0}(z)&\to&\left(\frac{2i}{\pi}\right)\Log(z)\label{eq:hank2}\\
H^{(1)}_{\nu}(z)&\to&-\left(\frac{i}{\pi}\right)\Gamma(\nu)e^{-\nu\Log(z/2)}, \qquad \Re(\nu)>0\label{eq:hank3}\\
H^{(1)}_{i\nu}(z)&\to&A_{\nu}e^{i\nu\Log(z/2)}+B_{\nu}e^{-i\nu\Log(z/2)}, \qquad \nu\in\mathbb{R}, \nu\ne0.\label{eq:hank4}
\eea
where
\be
A_{\nu}=\frac{1+\coth(\pi\nu)}{\Gamma(1+i\nu)}, \qquad
B_{\nu}=-\frac{\text{csch}(\pi\nu)}{\Gamma(1-i\nu)}.
\ee
Since our goal is to evaluate \eqref{eq:poincareratio2}, we are only interested in positive values of $z$, for which $\Log(z)=\ln(x)$. 
For finite $\epsilon$, as can be seen from \eqref{eq:hank1}, both integrals in the numerator and denominator of $F(\epsilon,x_m)$ converge as $x_m\to\infty$.
Moreover, \eqref{eq:hank2}$-$\eqref{eq:hank4} show that both integrals diverge in the limit $\epsilon\to0$, which means we can let $x_m=\infty$ and only concern ourselves with the behaviour of the integrands close to zero. Doing so, \eqref{eq:hank2} and \eqref{eq:hank3} imply
\be
\lim_{\substack{\epsilon\to0 \\ x_m\to\infty}}F(\epsilon,x_m)=-1 \qquad\text{for}\qquad
\nu\ge0.
\ee
Similarly, \eqref{eq:hank3} implies
\bea
\lim_{\substack{\epsilon\to0 \\ x_m\to\infty}}F(\epsilon,x_m)=\frac{2A_{\nu}B_{\nu}}{|A_{\nu}|^2+|B_{\nu}|^2}=-\sech(\pi|\nu|)\qquad\text{for} \qquad
\nu=i|\nu|, \nu\ne0.
\eea
To derive this last equality, we have used the following properties of the Gamma function (see $5.5.1$, $5.5.3$ and $5.4.3$ of \cite{Olver:2010:NHM:1830479}):
\bea
\Gamma(z+1)&=&z\Gamma(z)\label{gamma1}\\
\Gamma(z)\Gamma(1-z)&=&\frac{\pi}{\sin(\pi z)}\qquad z\ne0, \pm1,\pm2,\dots\label{gamma2}\\
|\Gamma(iy)|&=&\sqrt{\frac{\pi}{y\sinh(\pi y)}}\qquad y\in\mathbb{R}. \label{gamma3}
\eea
It then follows that
\be
\Gamma(1+i|\nu|)\Gamma(1-i|\nu|)=(i|\nu|)\Gamma(i|\nu|)\Gamma(1-i|\nu|)=i|\nu|\pi/\sin(i\pi|\nu|)=\pi|\nu|/\sinh(\pi|\nu|)
\ee
and
\be
|\Gamma(1\pm i|\nu|)|=|\pm i|\nu|\Gamma(\pm i |\nu|)|=\sqrt{\pi|\nu|/\sinh(\pi|\nu|)}.
\ee
Using these expressions we obtain
\bea
\frac{2A_{\nu}B_{\nu}}{|A_{\nu}|^2+|B_{\nu}|^2}=\frac{-2[1+\coth(\pi|\nu|)]\text{csch}(\pi|\nu|)}{[1+\coth(\pi|\nu|)]^2+\text{csch}^2(\pi|\nu|)}
=-\sech(\pi|\nu|).\notag
\eea
Figure~\ref{figrk1} provides numerical evidence for these calculations. We have computed $|F(\epsilon,x_m)|$ numerically and plotted its behaviour as a function of $\epsilon$.
These results are consistent with the analytical arguments provided above. 

Summarising our results:
\be
r_{k}=
\begin{dcases} 
e^{i\pi\left(\nu-\frac{d}{2}\right)} & \text{if $m\le m_*$,}
\\
e^{-i\pi\frac{d}2}\sech(\pi|\nu|) & \text{if $m\ge m_*$.}
\end{dcases}
\ee

\subsection{Global chart}
\label{app:global}
Here we shall evaluate \eqref{eq:globalratio}:
\be
r_L:=(-1)^L\frac{\langle \overline{y}^{E}_L,y^{E}_L\rangle_{t}}{\langle y^{E}_L,y^{E}_L\rangle_{t}}.
\ee
It follows from \eqref{minusCC} that
\bea
\langle y^E_L,y^E_L \rangle_t&=&2\int_0^T|y^E_L(t)|^2l^d\cosh^d(t/l)dt\\
\langle \overline{y}^E_L,y^E_L \rangle_t&=2&\int_0^T\text{Re}\left[y^E_L(t)^2\right]l^d\cosh^d(t/l)dt.
\eea
Changing integration variables to $z(t)=1+e^{2t/l}$, these integrals become
\be
\langle \overline{y}^E_L,y^E_L \rangle_t=\frac{l^{d+1}}{2^{2a}}\text{Re}\left\{\int_{2}^{z_T}\mathcal{N}_L^2I_1(z)dz\right\}, \qquad
\langle y^E_L,y^E_L \rangle_t=\frac{l^{d+1}}{2^{2a}}\int_{2}^{z_T}|\mathcal{N}_L|^2I_2(z)dz,
\ee
where
\be
I_1(z):=\frac{1}{\mathcal{N}_L^2}v_L(z)^2(z-1)^{\nu-1}z^{2a}, \qquad
I_2(z):=\frac{1}{|\mathcal{N}_L|^2}|v_L(z)|^2(z-1)^{\text{Re}(\nu)-1}z^{2a}.
\ee
Here $z_T=z(T)$ and all other quantities have been defined in Section \ref{AppEuc}. With these definitions:
\be
\lim_{z_T\to\infty}\frac{\langle \overline{y}^{E}_L,y^{E}_L\rangle_{t}}{\langle y^{E}_L,y^{E}_L\rangle_{t}}=\lim_{z_T\to\infty}\frac{\text{Re}\left\{\mathcal{N}_L^2\int_{2}^{z_T}I_1(z)dz\right\}}
{|\mathcal{N}_L|^2\int_{2}^{z_T}I_2(z)dz}.
\ee
The lower limit of these integrals is completely well-behaved, but they diverge in the limit where $z_T\to\infty$. Therefore, it suffices to study the integrands in this limit only.
Using the asymptotic behaviour of the hypergeometric
function given in \eqref{FinInf}, it can be checked that \textbf{when $\nu$ is real}: (see Section \ref{AppEuc} for definition of $\gamma$ and $\xi$)
\be
I_1(z)\xrightarrow{z\to\infty}\gamma^2 e^{-2i\pi a}z^{\nu-1}, \qquad
I_2(z)\xrightarrow{z\to\infty}|\gamma|^2 z^{\nu-1}.
\ee
Given that both quantities have the same scaling with $z$ in this limit, their ratio must converge to a constant when $z_T\to\infty$: 
\be
\lim_{z_T\to\infty}\frac{\langle \overline{y}^{E}_L,y^{E}_L\rangle_{t}}{\langle y^{E}_L,y^{E}_L\rangle_{t}}
=\frac{\text{Re}\left\{|\mathcal{N}_L|^2e^{2i\Theta}e^{-2i\pi a}\right\}}{|\mathcal{N}_L|^2}=\cos[\pi(\nu-a)],
\ee
where $\Theta=\frac{\pi}{2}[a+\text{Re}(\nu)]$.
Here we have used the fact that $\gamma=\overline{\gamma}$ when $\nu$ is real.
\textbf{When $\nu$ is imaginary}, it follows from \eqref{FinInf} that
\bea
I_1(z)&\xrightarrow{z\to\infty}&e^{-2i\pi a}z^{-1}\left[2\gamma\xi e^{\pi|\nu|}+\gamma^2e^{i|\nu|\ln(z)}+\xi^2e^{\pi|\nu|}e^{-i|\nu|\ln z}\right] \\
I_2(z)&\xrightarrow{z\to\infty}&z^{-1}\left[|\gamma|^2+|\xi|^2e^{2\pi|\nu|}+\gamma\overline{\xi} e^{i|\nu|\ln(z)}e^{\pi|\nu|}+\overline{\gamma}\xi e^{-i|\nu|\ln(z)}e^{\pi|\nu|}\right].
\eea
Again, since both quantities have the same scaling with $z$ in this limit, the ratio of their integrals converges to a constant as $z_T\to\infty$:
\bea
\lim_{z_T\to\infty}\frac{\langle \overline{y}^{E}_L,y^{E}_L\rangle_{t}}{\langle y^{E}_L,y^{E}_L\rangle_{t}}
&=&\frac{\text{Re}\left\{2\gamma\xi|\mathcal{N}_L|^2e^{2i\Theta}e^{-2i\pi a}\right\}}{|\mathcal{N}_L|^2(|\gamma|^2+|\xi|^2e^{2\pi\nu})}\\
&=&\cos(\pi a)\sech(\pi|\nu|),
\eea
having used the fact that $\gamma=\overline{\xi}$ when $\nu$ is imaginary. Notice that
\bea
\cos\left[\pi(\nu-a)\right]&=&\cos(\pi L+\pi d/2-\pi\nu)=(-1)^L\cos(\pi d/2-\pi\nu)\\
&=&(-1)^L\cos(\pi D/2-\pi\nu-\pi/2)=(-1)^L\sin(\pi D/2-\pi\nu).
\eea
Similarly, $\cos(\pi a)=(-1)^L\sin(\pi D/2)$.
Summarising our results:
\be
r_L=
\begin{dcases} 
\sin\frac{D}{2}\pi\,\textrm{sech}\,\pi|\nu| & \text{if $m\geq m_*$,}
\\
\sin\left[\left(\frac{D}{2}-|\nu|\right)\pi\right] &\text{if $0<m\le m_*$.}
\end{dcases}
\ee

\section{\label{app:sprinkling}Sprinkling into a diamond in $dS^2$}

To produce a sprinkling $\mathcal C_{M}$ into a causal diamond $M$ in $dS^2$, we need to pick a coordinate chart. The cosmological coordinates $x_P^\mu$ defined in~\eqref{eq:poincareCoord} are well suited because they have a conformally flat metric, which makes it particularly simple to compute the causal relation between points, given their coordinate values. Even though this chart only covers half of de Sitter space, there is no loss of generality because the symmetries of de Sitter space imply that any causal diamond can be isometrically mapped to a causal diamond entirely contained in the Poincar\'e patch.\\

Let $M$ be a causal diamond between two points $p,q\in dS^2_P$ such that $p\prec q$. Denote the (timelike) geodesic distance between $p$ and $q$ by $\tau$. Since any two causal diamonds with the same value of $\tau$ are isometric, we choose $x^\mu_p=(\eta_\tau,0)$ and $x^\mu_q=(\ell^2/\eta_\tau,0)$ with
\be
\eta_\tau=-\ell e^{\tau/2\ell}<-\ell,
\ee 
without loss of generality. To obtain a sprinkling $\mathcal C_ M$ into $M$ we first generate a uniform Poisson distribution of $N$ points in the square $[0,1]^2$ using a Mersenne Twister algorithm~\cite{matsumoto1998mersenne}. We use Cartesian coordinates $y_1,y_2$ on $[0,1]^2$ and find an embedding $\varphi:[0,1]^2\rightarrow R$, which for any subset $A \subset [0,1]^2$ satisfies
\be
V_{M}\int_A dy_1dy_2=\int_{\varphi(A)}d^2x\sqrt{-g}.
\label{eq:conditiononf}
\ee
The factor $V_{M}$ on the left hand side guarantees that the embedding scales the volume correctly. Its value for the causal diamond of length $\tau$ is
\be
\bal
V_{M}
=4\ell^2\ln\cosh\frac{\tau}{2\ell}.
\eal
\ee
By inspection it can be shown that the embedding $\varphi:(y_1,y_2)\rightarrow(\eta,x)$ defined by~\cite{Schmitzer:2010to}
\be
\bal
\eta&=\frac{-\ell e^{\tau/2\ell}}{1+y_1(e^{\tau/\ell}-1)},\\
r &=(1-2y_2)\sinh\frac{\tau}{2\ell},
\eal
\ee
satisfies the above condition~\eqref{eq:conditiononf}. By keeping only such points for which $|x|<\mathrm{min}(\eta_\tau-\eta,\eta-\ell^2/\eta_\tau)$ and recording the causal relations among them, we obtain a sprinkling $\mathcal C_M$ into $M$. Note that, as explained above, we also calculate the geodesic distance between any two points using the metric on the manifold, even though this data is not explicitly part of $\mathcal C_M$.\\

\bibliographystyle{jhep}
\bibliography{desitterpaper_v10}

\end{document}